\documentclass[12pt,twoside]{article}

\usepackage{latexsym}
\usepackage{times}

\input{psfig.sty}

\begin{document}

\newcommand{\findfigure}[1]{figures/#1.eps}

\newcommand{\comment}[1]{}
\newcommand{\enote}[1]{{\bf [Note: #1]}}

\setlength{\parskip}{1ex}
\setlength{\parindent}{0em}

\newtheorem{lemma}{Lemma}
\newtheorem{theorem}{Theorem}
\newtheorem{proposition}{Proposition}
\newtheorem{corollary}{Corollary}

\newcommand{\bnf}{\mathrel{::=}}
\newcommand{\kw}[1]{{\bf #1}}
\newcommand{\va}[1]{{\rm {\it #1}}}
\newcommand{\mva}[1]{\mbox{\va{#1}}}
\newcommand{\syntaxform}[1]{\mva{#1}}
\newcommand{\formula}{\syntaxform{formula}}
\newcommand{\pformula}{\syntaxform{p-formula}}
\newcommand{\gformula}{\syntaxform{g-formula}}
\newcommand{\sterm}{\syntaxform{term}}
\newcommand{\pterm}{\syntaxform{p-term}}
\newcommand{\gterm}{\syntaxform{g-term}}

\newcommand{\ITE}{\mva{ITE}}
\newcommand{\ite}{\mva{ite}}

\newcommand{\true}{\kw{true}}
\newcommand{\false}{\kw{false}}

\newcommand{\compare}[2]{#1\!=\!#2} 
\newcommand{\vbar}{\mathbin{|}}
\newcommand{\lvbar}{|\;}

\newcommand{\ftop}{F_{\mbox{top}}}
\newcommand{\formset}{\Phi}
\newcommand{\termset}{\Theta}
\newcommand{\nformset}{\formset^{-}}
\newcommand{\ntermset}{\termset^{-}}
\newcommand{\fsym}{{\cal F}}
\newcommand{\fsymp}{\fsym_{p}}
\newcommand{\fsymg}{\fsym_{g}}

\newcommand{\apply}[2]{#1(#2)}
\newcommand{\Apply}[2]{#1\left(#2\right)}

\newcommand{\eval}[2]{#1[#2]}
\newcommand{\Eval}[2]{#1\left[#2\right]}

\newcommand{\order}{{\it ord}}
\newcommand{\aorder}[1]{\order(#1)}
\newcommand{\domain}{{\cal D}}
\newcommand{\domainp}{{\cal D}'}
\newcommand{\domainh}{\hat{{\cal D}}}
\newcommand{\Dp}{\domain_{p}}
\newcommand{\Dg}{\domain_{g}}

\newcommand{\interp}{I}
\newcommand{\interpp}{I'}
\newcommand{\interps}{I^{*}}

\newcommand{\jinterp}{J}
\newcommand{\jinterpp}{J'}
\newcommand{\jinterph}{\hat{J}}
\newcommand{\jinterps}{J^{*}}
\newcommand{\interpt}{\tilde{I}}
\newcommand{\interpi}{\interp_{i}}
\newcommand{\interpimm}{\interp_{i-1}}

\newcommand{\ainterp}[1]{\apply{\interp}{#1}}
\newcommand{\ainterpp}[1]{\apply{\interpp}{#1}}
\newcommand{\ainterps}[1]{\apply{\interps}{#1}}
\newcommand{\ajinterp}[1]{\apply{\jinterp}{#1}}
\newcommand{\ajinterpp}[1]{\apply{\jinterpp}{#1}}
\newcommand{\ajinterph}[1]{\apply{\jinterph}{#1}}
\newcommand{\ajinterps}[1]{\apply{\jinterps}{#1}}
\newcommand{\ainterpt}[1]{\apply{\interpt}{#1}}
\newcommand{\ainterpi}[1]{\apply{\interpi}{#1}}
\newcommand{\ainterpimm}[1]{\apply{\interpimm}{#1}}

\newcommand{\einterp}[1]{\eval{\interp}{#1}}
\newcommand{\einterpp}[1]{\eval{\interpp}{#1}}
\newcommand{\einterps}[1]{\eval{\interps}{#1}}
\newcommand{\einterpt}[1]{\eval{\interpt}{#1}}
\newcommand{\ejinterp}[1]{\eval{\jinterp}{#1}}
\newcommand{\ejinterpp}[1]{\eval{\jinterpp}{#1}}
\newcommand{\ejinterph}[1]{\eval{\jinterph}{#1}}
\newcommand{\ejinterps}[1]{\eval{\jinterps}{#1}}
\newcommand{\einterpi}[1]{\eval{\interpi}{#1}}
\newcommand{\einterpimm}[1]{\eval{\interpimm}{#1}}

\newcommand{\tF}{\tilde{F}}

\newcommand{\hF}{\hat{F}}
\newcommand{\hG}{\hat{G}}
\newcommand{\hS}{\hat{S}}
\newcommand{\hT}{\hat{T}}
\newcommand{\hU}{\hat{U}}
\newcommand{\hE}{\hat{E}}
\newcommand{\hD}{\hat{D}}

\newcommand{\sF}{F^{*}}
\newcommand{\egF}{F_{\rm eg}}
\newcommand{\segF}{\egF^{*}}

\newcommand{\Sym}{\Sigma}
\newcommand{\Symp}{\Sym_{p}}
\newcommand{\Symps}{\Sym^{*}_{p}}
\newcommand{\Symgs}{\Sym^{*}_{g}}

\newcommand{\vf}{{\it vf}}
\newcommand{\vg}{{\it vg}}
\newcommand{\vh}{{\it vh}}

\newcommand{\depend}[1]{{\it depend}(#1)}
\newcommand{\encodef}[1]{{\it encf}(#1)}
\newcommand{\encodet}[2]{{\it enct}_{#2}(#1)}
\newcommand{\select}[2]{{\it sel}_{#1}(#2)}
\newcommand{\forder}[1]{o_f(#1)}
\newcommand{\lm}{{\it lm}}
\newcommand{\lmr}[3]{\lm_{#1}(#2,#3)}
\newcommand{\lma}[2]{\lm_{#1}(#2)}
\newcommand{\edgeval}[2]{e_{[#1,#2]}}

\newcommand{\remap}{\tau}

\newcommand{\fsat}{F_{\mbox{sat}}}

\newcommand{\depth}{{\it depth}}
\newcommand{\kmax}{k_{\max}}

\newcommand{\termud}[2]{T^{(#1)}_{#2}}
\newcommand{\stermud}[2]{S^{(#1)}_{#2}}
\newcommand{\gformu}[1]{G^{(#1)}}
\newcommand{\expru}[1]{E^{(#1)}}
\newcommand{\dexpru}[1]{{D}^{(#1)}}
\newcommand{\ordth}[1]{#1^{\mbox{\small th}}}
\newcommand{\ordst}[1]{#1^{\mbox{\small st}}}
\newcommand{\onetoone}{1--1}

\newcommand{\qed}{\mbox{} $\Box$}
\newenvironment{proof}{{\it Proof:}}{\qed}

\newcommand{\tset}{{\cal T}}
\newcommand{\equivi}{\mathrel{\approx_{\interp}}}
\newcommand{\equivip}{\mathrel{\approx_{\interpp}}}
\newcommand{\nequivi}{\mathrel{\not\approx_{\interp}}}
\newcommand{\nequivip}{\mathrel{\not\approx_{\interpp}}}

\title{Processor Verification Using
Efficient Reductions \\ of the Logic of Uninterpreted Functions\\
to Propositional Logic\footnote{A preliminary version of this paper was
published as \cite{bryant-cav99}}.}

\author{\begin{small}\begin{tabular}{ccc}
Randal E. Bryant\thanks{Supported by grants from Intel, Motorola, and Fujitsu.} & Steven German & Miroslav N. Velev\thanks{Supported by SRC Contract 98-DC-068.} \\
{\small\tt Randy.Bryant@cs.cmu.edu} & {\small\tt german@watson.ibm.com} &
{\small\tt mvelev@ece.cmu.edu} \\
Computer Science & Watson Research Center & Elec.~\& Comp.~Engineering \\
Carnegie Mellon University & IBM & Carnegie Mellon University\\
Pittsburgh, PA & Yorktown Heights, NY & Pittsburgh, PA\\
\end{tabular}\end{small}}

\maketitle

\begin{abstract}

The logic of equality with uninterpreted functions (EUF) provides a
means of abstracting the manipulation of data by a processor when
verifying the correctness of its control logic.  By reducing formulas
in this logic to propositional formulas, we can apply Boolean methods
such as Ordered Binary Decision Diagrams (BDDs) and Boolean
satisfiability checkers to perform the verification.

We can exploit characteristics of the formulas describing the
verification conditions to greatly simplify the propositional formulas
generated.  We identify a class of terms we call ``p-terms'' for which
equality comparisons can only be used in monotonically positive
formulas.  By applying suitable abstractions to the hardware model, we
can express the functionality of data values and instruction addresses
flowing through an instruction pipeline with p-terms.  A decision
procedure can exploit the restricted uses of p-terms by considering
only ``maximally diverse'' interpretations of the associated function
symbols, where every function application yields a different value
except when constrained by functional consistency.

We present two methods to translate formulas in EUF into propositional
logic.  The first interprets the formula over a domain of fixed-length
bit vectors and uses vectors of propositional variables to encode
domain variables.  The second generates formulas encoding the
conditions under which pairs of terms have equal valuations,
introducing propositional variables to encode the equality relations
between pairs of terms.  Both of these approaches can exploit maximal
diversity to greatly reduce the number of propositional variables that
need to be introduced and to reduce the overall formula sizes.

We present experimental results demonstrating the efficiency of this
approach when verifying pipelined processors using the method proposed
by Burch and Dill.  Exploiting positive equality allows us to overcome
the exponential blow-up experienced previously \cite{velev-fmcad98}
when verifying microprocessors with load, store, and branch
instructions.

\end{abstract}

\section{Introduction}

For automatically reasoning about pipelined processors, Burch and Dill
demonstrated the value of using propositional logic, extended with
uninterpreted functions, uninterpreted predicates, and the testing of
equality \cite{burch-cav94}.  Their approach involves abstracting
the data path as a collection of registers and memories storing data,
units such as ALUs operating on the data, and various connections and
multiplexors providing methods for data to be transferred and
selected.  The initial state of each register is represented by a
domain variable indicating an arbitrary data value.  The
operation of units that transform data is abstracted as blocks
computing functions with no specified properties other than
functional consistency, i.e., that applications of a function to
equal arguments yield equal results: $x = y \Rightarrow f(x) =
f(y)$.  The state of a register at any point in the computation can be
represented by a symbolic term, an expression consisting of a
combination of domain variables, function and predicate applications,
and Boolean operations.
Verifying that a pipelined processor has behavior matching that of an
unpipelined instruction set reference model can be performed by
constructing a formula in this logic that compares for equality the
terms describing the results produced by the two models and then
proving the validity of this formula.  

In their 1994 paper, Burch and Dill also described the implementation
of a decision procedure for this logic based on theorem proving search
methods.  Their procedure builds on ones originally described by
Shostak \cite{shostak-jacm79} and by Nelson and Oppen
\cite{nelson-jacm80}, using combinatorial search
coupled with algorithms for maintaining a partitioning of the terms
into equivalence classes based on the equalities that hold at a given
step of the search.  More details of their decision procedure are
given in \cite{jones-iccad95}.  

Burch and Dill's work has generated considerable interest in the use
of uninterpreted functions to abstract data operations in processor
verification.  A common theme has been to adopt Boolean methods,
either to allow integration of uninterpreted functions into symbolic
model checkers \cite{damm-concur98,berezin-fmcad98}, or to allow the
use of Binary Decision Diagrams (BDDs) \cite{bryant-ieeetc86} 
in the decision procedure
\cite{hojati-iwls97,goel-cav98,velev-fmcad98}.  Boolean methods allow
a more direct modeling of the control logic of hardware designs and
thus can be applied to actual processor designs rather than highly
abstracted models.  In addition to BDD-based decision procedures,
Boolean methods could use some of the recently developed
satisfiability procedures for propositional logic.  In principle,
Boolean methods could outperform decision procedures based on theorem
proving search methods, especially when verifying processors with more
complex control logic, e.g., due to superscalar or out-of-order
operation.

Boolean methods can be used to decide the validity of a formula
containing terms and uninterpreted functions by interpreting the
formula over a domain of fixed-length bit vectors.  Such an approach
exploits the property that a given formula contains a limited number
of function applications and therefore can be proved to be universally
valid by considering its interpretation over a sufficiently large, but
finite domain \cite{ackermann-54}.  If a formula contains a total of
$m$ function applications, then the set of all bit vectors of length
$k$ forms an adequate domain for $k \geq \log_2 m$.  The formula to be
verified can be translated into one in propositional logic, using
vectors of propositional variables to encode the possible values
generated by function applications \cite{hojati-iwls97}.  Our
implementation of such an approach \cite{velev-fmcad98} as part of a
BDD-based symbolic simulation system was successful at verifying
simple pipelined data paths.  We found, however, that the
computational resources grew exponentially as we increased the
pipeline depth.  Modeling the interactions between successive
instructions flowing through the pipeline, as well as the functional
consistency of the ALU results, precludes having an ordering of the
variables encoding term values that yields compact BDDs.  Similarly,
we found that extending the data path to a complete processor by
adding either load and store instructions or instruction fetch logic
supporting jumps and conditional branches led to impossible BDD
variable ordering requirements.

Goel {\em et al.}\  \cite{goel-cav98} present an alternate approach to
using BDDs to decide the validity of formulas in the logic of equality
with uninterpreted functions.  In their formulation they introduce a
propositional variable $e_{i,j}$ for each pair of function application
terms $T_i$ and $T_j$, expressing the conditions under which the two
terms are equal.  They add constraints expressing both functional
consistency and the transitivity of equality among the terms.  Their
experimental results were also somewhat disappointing.  For all
previous methods of reducing EUF to propositional logic, Boolean
methods have not lived up to their promise of outperforming ones based
on theorem proving search.

In this paper, we show that the characteristics of the formulas
generated when modeling processor pipelines can be exploited to
greatly reduce the number of propositional variables that are
introduced when translating the formula into propositional logic.  We
distinguish a class of terms we call {\em p-terms} for which equality
comparisons can be used only in monotonically positive formulas.  Such
formulas are suitable for describing the top-level correctness
condition, but not for modeling any control decisions in the hardware.
By applying suitable abstractions to the hardware model, we can
express the functionality of data values and instruction addresses
with p-terms.

A decision procedure can exploit the restricted uses of p-terms by
considering only ``maximally diverse'' interpretations of the
associated ``p-function'' symbols, where every function application
yields a different value except when constrained by functional
consistency.  We present a method of transforming a formula containing
function applications into one containing only domain variables that
differs from the commonly-used method described by Ackermann
\cite{ackermann-54}.  Our method allows a translation into
propositional logic that uses vectors with fixed bit patterns rather
than propositional variables to encode domain variables introduced
while eliminating p-function applications.  This reduction in
propositional variables greatly simplifies the BDDs generated when
checking tautology, often avoiding the exponential blow-up experienced
by other procedures.  Alternatively, we can use a encoding scheme
similar to Goel {\em et al.}\  \cite{goel-cav98}, but with many of the
$e_{i,j}$ values set to $\false$ rather than to Boolean variables.


Others have recognized the value of restricting the testing of
equality when modeling the flow of data in pipelines.  Berezin {\em et
al.}\  \cite{berezin-fmcad98} generate a model of an execution unit
suitable for symbolic model checking in which the data values and
operations are kept abstract.  In our terminology, their functional
terms are all p-terms.  They use fixed bit
patterns to represent the initial states of registers, much as we
replace p-term domain variables by fixed bit patterns.  To model the
outcome of each program operation, they generate an entry in a
``reference file'' and refer to the result by a pointer to this file.
These pointers are similar to the bit patterns we generate to
denote the p-function application outcomes.
This paper provides an alternate, and somewhat more general
view of the efficiency gains allowed by p-terms.

Damm {\em et al.}\  consider an even more restricted logic such that in
the terms describing the computed result, no function symbol is
applied to a term that already contains the same symbol.  As a
consequence, they can guarantee that an equality between two terms
holds universally if it holds holds over the domain $\{0,1\}$ and with
function symbols having four possible interpretations: constant
functions 0 or 1, and projection functions selecting the first or
second argument.  They can therefore argue that verifying an execution
unit in which the data path width is reduced to a single bit and in
which the functional units implement only four functions suffices to
prove its correctness for all possible widths and functionalities.
Their work imposes far greater restrictions than we place on p-terms,
but it allows them to bound the domain that must be considered to
determine universal validity independently from the formula size.

In comparison to both of these other efforts, we maintain the full
generality of the unrestricted terms of Burch and Dill while
exploiting the efficiency gains possible with p-terms.  In our
processor model, we can abstract register identifiers as unrestricted
terms, while modeling program data and instruction data as p-terms.
As a result, our verifications cover designs with arbitrarily many
registers.  In contrast, both \cite{berezin-fmcad98} and
\cite{damm-concur98} used bit encodings of register identifiers and
were unable to scale their verifications to a realistic number of
registers.

In a recent paper, Pnueli, {\em et al.}\ \cite{pnueli-cav99} also
propose a method to exploit the polarity of the equations in a formula
containing uninterpreted functions with equality.  They describe an
algorithm to generate a small domain for each domain variable such
that the universal validity of the formula can be determined by
considering only interpretations in which the variables range over
their restricted domains.  A key difference of their work is that they
examine the equation structure after replacing all function
application terms with domain variables and introducing functional
consistency constraints as described by Ackermann \cite{ackermann-54}.
These consistency constraints typically contain large numbers of
equations---far more than occur in the original formula---that mask
the original p-term structure.  As an example, comparing the top and
bottom parts of Figure \ref{ackermann-figure} illustrates the large
number of equations that may be generated when applying Ackermann's
method.  By contrast, our method is based on the original formula
structure.  In addition, we use a new method of replacing function
application terms with domain variables.  Our scheme allows us to
exploit maximal diversity by assigning fixed values to the domain
variables generated while expanding p-function application terms.
Quite possibly, a variant of their method could be used to generate a
small domain for each of the other variables in the formula.

The remainder of the paper is organized as follows.  We define the
syntax and semantics of our logic by extending that of Burch and
Dill's.  We describe a simple procedure for automatically converting
a formula from Burch and Dill's logic to ours.
We prove our central result concerning the need to
consider only maximally diverse interpretations when deciding the validity of
formulas in our logic.  As a first step in transforming our logic into
propositional logic, we describe a new method of eliminating function
application terms in a formula.  Building on this, we describe two
methods of translating formulas into propositional logic and show how
these methods can exploit the properties of p-terms.  We discuss the
abstractions required to model processor pipelines in our logic.
Finally, we present experimental results showing our ability to verify
a simple, but complete pipelined processor.  More complete details on
an implementation that has successfully verified several superscalar
processor designs are presented in \cite{velev-charme99}.

\section{Logic of Equality with Uninterpreted Functions (EUF)}

\begin{figure}
\begin{eqnarray*}
\sterm & \bnf  & \ITE(\formula, \sterm, \sterm) \\
&& \lvbar \syntaxform{function-symbol}(\sterm, \ldots, \sterm) \\
\\
\formula & \bnf  & \true \vbar \false \vbar \neg \formula \\
&& \lvbar (\formula \land \formula) \vbar (\formula \lor \formula) \\
&& \lvbar  (\compare{\sterm}{\sterm})  \\
&& \lvbar \syntaxform{predicate-symbol}(\sterm, \ldots, \sterm) 
\end{eqnarray*}
\caption{Syntax Rules for the Logic of Equality with
Uninterpreted Functions (EUF)}
\label{euf-figure}
\end{figure}

The logic of Equality with Uninterpreted Functions (EUF) presented by
Burch and Dill \cite{burch-cav94} can be expressed by the syntax given
in Figure \ref{euf-figure}.  In this logic, {\em formulas} have truth
values while {\em terms} have values from some arbitrary domain.
Terms are formed by application of uninterpreted function symbols and
by applications of the $\ITE$ (for ``if-then-else'') operator.  The
$\ITE$ operator chooses between two terms based on a Boolean control
value, i.e., $\ITE(\true, x_1, x_2)$ yields $x_1$ while $\ITE(\false,
x_1, x_2)$ yields $x_2$.  Formulas are formed by comparing two terms
with equality, by applying an uninterpreted predicate symbol to a list
of terms, and by combining formulas using Boolean connectives.  A
formula expressing equality between two terms is called an {\em
equation}.  We use {\em expression} to refer to either a term or a
formula.

Every function symbol $f$ has an associated {\em order}, denoted
$\aorder{f}$, indicating the number of terms it takes as arguments.
Function symbols of order zero are referred to as {\em domain
variables}.  We use the shortened form $v$ rather than $v()$ to denote
an instance of a domain variable.  Similarly, every predicate $p$ has an
associated order $\aorder{p}$.  Predicates of order zero are referred
to as {\em propositional variables}, and can be written $a$ rather
than $a()$.

\begin{table}
\begin{center}
\begin{tabular}{|c|c|}
\hline
Form $E$ & Valuation $\einterp{E}$ \\
\hline
\true & \true \\
\false & \false \\
$\neg F$ & $\neg \einterp{F}$ \\
$F_1 \land F_2$ & $\einterp{F_1} \land \einterp{F_2}$ \\
$p(T_1, \ldots, T_k)$ & $\ainterp{p}(\einterp{T_1}, \ldots, \einterp{T_k})$ \\
$\compare{T_1}{T_2}$ & $\compare{\einterp{T_1}}{\einterp{T_2}}$ \\
\hline
$\ITE(F, T_1, T_2)$ & $\ITE(\einterp{F}, \einterp{T_1}, \einterp{T_2})$ \\
$f(T_1, \ldots, T_k)$ & $\ainterp{f}(\einterp{T_1}, \ldots, \einterp{T_k})$ \\
\hline
\end{tabular}
\end{center}
\caption{Evaluation of EUF Formulas and Terms}
\label{interp-table}
\end{table}

The truth of a formula is defined relative to a nonempty domain $\domain$ of
values and an interpretation $\interp$ of the function and predicate
symbols.  Interpretation $\interp$ assigns to each function symbol of
order $k$ a function from $\domain^k$ to $\domain$, and to each
predicate symbol of order $k$ a function from $\domain^k$ to $\{\true,
\false \}$.  For the special case of order 0 symbols, i.e., domain
(respectively, propositional) variables, the interpretation assigns an element of $\domain$ (resp., $\{\true, \false\}$.)  Given
an interpretation $\interp$ of the function and predicate symbols and
an expression $E$, we can define the {\em valuation} of $E$ under $\interp$,
denoted $\einterp{E}$, according to its syntactic structure.  The
valuation is defined recursively, as shown in Table
\ref{interp-table}.  $\einterp{E}$ will be an element of the domain
when $E$ is a term, and a truth value when $E$ is a formula.

A formula $F$ is said to be {\em true under interpretation $\interp$}
when $\einterp{F} = \true$.  It is said to be {\em valid over domain
$\domain$} when it is true over domain $\domain$ for all
interpretations of the symbols in $F$.  $F$ is said to be {\em
universally valid} when it is valid over all domains.  A basic
property of validity is that a given formula is valid over a domain
$\domain$ iff it is valid over all domains having the same cardinality
as $\domain$.  This follows from the fact that a given formula has the
same truth value in any two isomorphic interpretations of the symbols
in the formula.  Another property of the logic, which can be readily
shown, is that if $F$ is valid over a suitably large domain, then it
is universally valid \cite{ackermann-54}.  In particular, it suffices
to have a domain as large as the number of syntactically distinct
function application terms occurring in $F$.  We are interested in
decision procedures that determine whether or not a formula is
universally valid; we will show how to do this by dynamically
constructing a sufficiently large domain as the formula is being
analyzed.

\section{Positive Equality with Uninterpreted Functions (PEUF)}
\begin{figure}
\begin{eqnarray*}
\gterm & \bnf  & \ITE(\gformula, \gterm, \gterm) \\
&& \lvbar \syntaxform{g-function-symbol}(\pterm, \ldots, \pterm) \\
\\
\pterm & \bnf  & \gterm \\
&& \lvbar \ITE(\gformula, \pterm, \pterm) \\ 
&& \lvbar \syntaxform{p-function-symbol}(\pterm, \ldots, \pterm) \\
\\
\gformula & \bnf  &  \true \vbar \false \vbar \neg \gformula \\ 
&& \lvbar (\gformula \land \gformula) \vbar (\gformula \lor \gformula) \\
&& \lvbar ( \compare{\gterm}{\gterm} ) \\ 
&& \lvbar \syntaxform{predicate-symbol}(\pterm, \ldots, \pterm) \\
\\
\pformula & \bnf  & \gformula \\
&& \lvbar (\pformula \land \pformula) \vbar (\pformula \lor \pformula) \\
&& \lvbar ( \compare{\pterm}{\pterm} )
\end{eqnarray*}
\caption{Syntax Rules for the Logic of Positive Equality with
Uninterpreted Functions (PEUF)}
\label{peuf-figure}
\end{figure}

We can improve the efficiency
of validity checking by treating positive and negative equations
differently when reducing EUF to propositional logic.  Informally, an
equation is positive if it does not appear negated in a formula.  In
particular, a positive equation cannot appear as the formula that
controls the value of an $\ITE$ term; such formulas are considered to
appear both positively and negatively.

\comment{
We will find it useful to
recognize sets of function application terms that appear only
positively in a sense defined below.
}

\comment{
We will show that validity checking can be made more efficient in some
cases if we divide the function symbols into two disjoint classes,
called {\em p-function} symbols and {\em g-function} symbols.  
We will define an extended logic with syntactic restrictions such that
an application of a p-function symbol cannot appear in a negative
equation, while g-function symbols have no such restriction.
}

\subsection{Syntax}

PEUF is an extended logic based on EUF; its syntax is shown in Figure
\ref{peuf-figure}.  The main idea is that there are two disjoint classes
of function symbols, called p-function symbols and g-function symbols,
and two classes of terms.

General terms, or {\em g-terms}, correspond to terms in EUF{}.
Syntactically, a g-term is a g-function application or an $\ITE$ term
in which the two result terms are hereditarily built from g-function
applications and $\ITE$s.

The new class of terms is called positive terms, or {\em p-terms}.
P-terms may not appear in negated equations, i.e., equations within
the scope of a logical negation.  Since p-terms can contain p-function
symbols, the syntax is restricted in a way that prevents p-terms from
appearing in negative equations.  When two p-terms are compared for
equality, the result is a special, restricted kind of formula called a
{\em p-formula}.

Note that our syntax allows any g-term to be ``promoted'' to a p-term.
Throughout the syntax definition, we require function and predicate
symbols to take p-terms as arguments.  However, since g-terms can be
promoted, the requirement to use p-terms as arguments does not
restrict the use of g-function symbols or g-terms.  In essence,
g-function symbols may be used as freely in our logic as in EUF, but the
p-function symbols are restricted.  To maintain the restriction on
p-function symbols, the syntax does not permit a p-term to be promoted
to a g-term.

A {\em g-formula} is a Boolean combination of equations on g-terms and
applications of predicate symbols.  G-formulas in our logic serve as
Boolean control expressions in $\ITE$ terms.  A g-formula can contain
negation, and $\ITE$ implicitly negates its Boolean control, so only
g-terms are allowed in equations in g-formulas.

Finally, the syntactic class {\em p-formula} is the class for which we
develop validity checking methods.
p-formulas are built up using only the monotonically
positive Boolean operations $\land$ and $\lor$.  P-formulas may not be placed
under a negation sign and cannot be used as the control for an $\ITE$
operation.
As described in later sections, our validity checking
methods will take advantage of the assumption that in p-formulas, 
the p-terms cannot appear in negative equations.

\begin{figure}
\centerline{\psfig{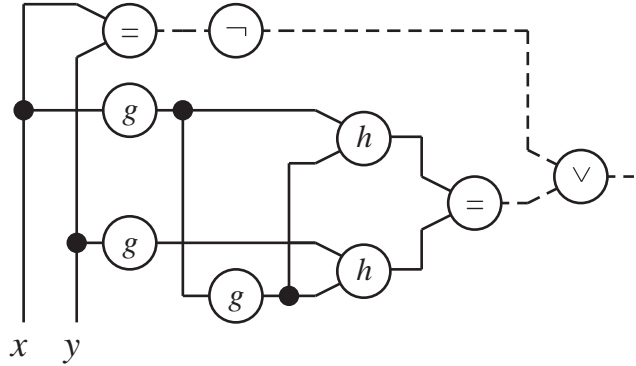}}
\caption{Schematic Representation of $\egF$.  Domain values are shown as solid lines, while truth values
are shown as dashed lines.}
\label{eg-start-figure}
\end{figure}

As a running example for this paper, we consider the formula
$\compare{x}{y} \Rightarrow \compare{h(g(x),g(g(x)))}{h(g(y),g(g(x)))}$,
which would
be transformed into a p-formula $\egF$ by eliminating the implication:
\begin{eqnarray}
\egF & = & 
\neg (\compare{x}{y}) \;\;\lor \;\;
\compare{h(g(x),g(g(x)))}{h(g(y),g(g(x)))}
\label{eg-formula}
\end{eqnarray}
Domain variables $x$ and $y$ must be g-function symbols so that we can
consider the equation $\compare{x}{y}$ to be a g-formula, and hence it
can be negated to give g-formula $\neg (\compare{x}{y})$.  We can
promote the g-terms $x$ and $y$ to p-terms, and we can consider
function symbols $g$ and $h$ to be p-function symbols, giving p-terms
$g(x)$, $g(y)$, $g(g(x))$, $h(g(x), g(g(x)))$, and $h(g(y), g(g(x)))$.
Thus, the equation $\compare{h(g(x), g(g(x)))}{h(g(y), g(g(x)))}$ is a
p-formula.  We form the disjunction of this p-formula with the
p-formula obtained by promoting $\neg (\compare{x}{y})$ giving
p-formula $\egF$.

Figure \ref{eg-start-figure} shows a
schematic representation of $\egF$, using drawing
conventions similar to those found in hardware designs.  That is, we
view domain variables as inputs (shown along bottom) to a network of
operators.  Domain values are denoted with solid lines, while truth
values are denoted with dashed lines.  The top-level formula then
becomes the network output, shown on the right.  The operators in the
network are shared whenever possible.  This representation is
isomorphic to the traditional directed acyclic graph (DAG)
representation of an expression, with maximal sharing of common
subexpressions.

\subsection{Extracting PEUF from EUF}

Observe that PEUF does not extend the expressive power of EUF---we
could translate any PEUF expression into EUF by considering both the
p-terms and g-terms to be terms and both the p-formulas and g-formulas
to be formulas.  Instead, the benefit of PEUF is that by
distinguishing some portion of a formula as satisfying a restricted
set of properties, we can radically reduce the number of different
interpretations we must consider when proving that a p-formula is
universally valid.

In fact, we can automatically extract the PEUF syntax from an EUF
formula by the following process, and hence our decision procedure can
be viewed as one that automatically exploits the polarity structure of
equations in an arbitrary EUF formula $\ftop$.  The main task is to classify
the function symbols as either p-function or g-function symbols.

We assume our EUF formula $\ftop$ is in {\em negation-normal} form,
meaning that the negation operation $\neg$ is applied only to
equations and predicate applications.  We can convert an arbitrary
formula into negation-normal form by applying the following syntactic
transformations:
\begin{eqnarray*}
\neg \true & \rightarrow & \false \\
\neg \false & \rightarrow & \true \\
\neg \neg F & \rightarrow & F \\
\neg (F_1 \land F_2) & \rightarrow & \neg F_1 \lor \neg F_2 \\
\neg (F_1 \lor F_2) & \rightarrow & \neg F_1 \land \neg F_2 \\
\end{eqnarray*}

To formalize the relationship between EUF expressions and PEUF
expressions, we introduce a {\em tree representation} of
EUF expressions.  The rules for the tree representation are as follows:

\begin{enumerate}
\item If $E$ is an EUF expression having no proper subexpressions
($\true$, $\false$, a domain variable, or a propositional variable),
then $E$ is represented by a tree consisting of a single node labelled
with $E$.

\item If $E$ is an EUF expression having $n$ proper subexpressions,
then $E$ is represented by a tree whose root node is labelled with
the main operator ($=$, $\ITE$, $\land$, $\lor$, $\neg$, predicate symbol,
function symbol).  Attached to the root node are $n$ subtrees, where
the $i$th subtree represents the $i$th proper subexpression.
\end{enumerate}

We define a {\em parsing} of an EUF expression as a PEUF
expression.  Let $t$ be a tree representing an EUF expression $E$.  A
parsing of $E$ as a PEUF expression is a function that assigns to each
node of $t$ a set of syntax classes in the formal syntax of PEUF, such
that the syntax rules of PEUF (Figure \ref{peuf-figure}) are
satisfied.
Note that this definition allows multiple syntax classes to be
assigned to a given tree node.  This multiplicity arises due to the
two syntax rules: $\pformula \bnf \gformula$, and $\pterm \bnf \gterm$.
That is, every tree node that can be classified as a g-formula
(respectively, g-term) can also be classified as a p-formula (resp.,
p-term).



We say there is a parsing of an EUF expression $E$ as a PEUF
expression of a given syntax class $\mva{cl}$, if there is a parsing
of a tree representing $E$ that satisfies the PEUF syntax rules, and
$\mva{cl}$ is in the set of syntax classes assigned to the root node
of the tree.

To state the main result of this section about parsing, we first
define several sets of expressions.  Let $\formset$ (respectively
$\termset$) be the set of all syntactically-distinct formulas (resp.,
terms) occurring in $\ftop$.  We
define the set $\nformset \subseteq \formset$ of negative formulas to
be the smallest set of formulas satisfying the following conditions:
\begin{enumerate}
\item For every formula $\neg F$ in $\formset$, formula $F$ is in $\nformset$.
\item For every term $\ITE(F, T_1, T_2)$ in $\termset$, formula $F$ is
in $\nformset$.
\item For every formula $F_1 \land F_2$ in $\nformset$, formulas $F_1$
and $F_2$ are in $\nformset$.
\item For every formula $F_1 \lor F_2$ in $\nformset$, formulas $F_1$
and $F_2$ are in $\nformset$.
\end{enumerate}

We define the set $\ntermset \subseteq \termset$ of negative terms to
be the smallest set of terms satisfying:
\begin{enumerate}
\item 
For every equation $\compare{T_1}{T_2}$ in $\nformset$, terms
$T_1$ and $T_2$ are in $\ntermset$.
\item For every term $\ITE(F, T_1, T_2)$ in $\ntermset$, terms
$T_1$ and $T_2$ are in $\ntermset$.
\end{enumerate}

Finally, we partition the set of all function symbols $\fsym$ into
disjoint sets $\fsymg$ and $\fsymp$ as follows.  If there is some term
in $\ntermset$ of the form $f(T_1, \ldots, T_k)$, then $f$ is in
$\fsymg$.  If there is no such term, then $f$ is in $\fsymp$.

\begin{theorem}
For any negation-normal EUF formula $\ftop$, there is 
a parsing of $\ftop$ as a PEUF p-formula such that each
function symbol in $\fsymg$ is a g-function symbol, and each function
symbol in $\fsymp$ is a p-function symbol.
\label{peuf-extraction-theorem}
\end{theorem}

\begin{proof}

For the remainder of this proof, we consider a fixed EUF formula
$\ftop$.  We will only consider a function to be a parsing if it is a
parsing when the set of g-function symbols is $\fsymg$ and the set of
p-function symbols is $\fsymp$.


We prove this theorem by induction on the syntactic structure of
$\ftop$.  Our induction hypothesis consists of four
assertions, two for terms and two for formulas:
\begin{enumerate}
\item For $T \in \termset$ such that $T \in \ntermset$ or
$T$ is a function application with a function symbol in $\fsymg$,
there is a parsing of $T$ as a g-term.

\item For $T \in \termset$, there is a parsing of $T$ as a p-term.

\item For $F \in \formset$ satisfying one of the following conditions:
\begin{enumerate}
\item $F$ is $\true$ or $\false$,
\item $F$ is a formula of the form $\neg F_1$,
\item $F$ is a predicate application,
\item $F$ is in $\nformset$,
\end{enumerate}
there is a parsing of $F$ as a g-formula.

\item For  $F \in \formset$, there is a parsing of $F$ as a p-formula.
\end{enumerate}

Recall that the syntax of PEUF allows any g-formula to be promoted to a
p-formula, and any g-term to be promoted to a p-term.  These promotion
rules will be used several times in the proof.

For the base cases, we consider expressions having no proper subexpressions:
\begin{enumerate}
\item For a domain variable $v$, if $v \in \ntermset$, then $v \in
\fsymg$, so there is a parsing of $v$ as a g-term and a parsing as a p-term.

\item For a domain variable $v \in \termset - \ntermset$, $v$ is in 
$\fsymp$, so there is a parsing of $v$ as a p-term.

\item EUF formulas $\true{}$ and $\false{}$ can be parsed as either
g-formulas or p-formulas.

\item For a propositional variable $p$, there is a parsing of $p$ as a
g-formula or as a p-formula.
\end{enumerate}

For the inductive argument, we prove the following cases for EUF
expressions, assuming that all proper subexpressions obey the
induction hypothesis.
\begin{enumerate}
\item Terms in $\termset$:
\begin{enumerate}
\item Consider $T \doteq \ITE(F, T_1, T_2)$.
If $T \in \ntermset$, then by definition, $F \in \nformset$ and
$T_1, T_2 \in \ntermset$.  Thus, by the inductive hypothesis, 
there are parsings of $F$ as a g-formula and of 
$T_1$ and $T_2$ as g-terms.  This means there is a parsing of $T$ as a
g-term.

If $T \in \termset$, then by the inductive hypothesis, there are
parsings of $F$ as a g-formula and of $T_1$ and $T_2$ as p-terms.  Thus
there is a parsing of $T$ as a p-term.

\item Consider $T \doteq f(T_1, \ldots, T_k)$.  By the inductive
hypothesis, there are parsings of $T_1, \ldots, T_k$ as p-terms.  When
$f \in \fsymg$, there are parsings of $T$ as a g-term and, by
promotion, as a p-term.  When $f \in \fsymp$, there is a parsing of
$T$ as a p-term.  Thus, there is a parsing of $T$ as a p-term in
either case.  In addition, when $T \in \ntermset$, we must have $f \in
\fsymg$, and hence there is also a parsing of $T$ as a g-term.
\end{enumerate}
\item Formulas in $\formset$:
\begin{enumerate}
\item Consider $F \doteq \neg F_1$.  We have $F_1 \in \nformset$, so
there is a parsing of $F_1$ as a g-formula.  Hence $F$ can be parsed as a
g-formula or a p-formula.

\item Consider $F \doteq F_1 \land F_2$.  If $F$ is in $\nformset$,
then $F_1, F_2$ are in $\nformset$, so $F_1, F_2$ can be parsed as g-formulas
and $F$ can be parsed as a g-formula or as a p-formula.

If $F$ is in $\formset$, then $F_1, F_2$ can be parsed as p-formulas,
so $F$ can be parsed as a p-formula.

\item Consider $F \doteq F_1 \lor F_2$.  Similar to previous case.

\item Consider $F \doteq \compare{T_1}{T_2}$.  If $F \in \nformset$, then $T_1,
T_2 \in \ntermset$ and hence $T_1$ and $T_2$ can be parsed as g-terms,
so $F$ can be parsed as a g-formula or as a p-formula.

If $F \in \formset$, then $T_1$ and $T_2$ can be parsed as p-terms, so $F$
can be parsed as a p-formula.

\item Consider $F \doteq p(T_1, \ldots, T_k)$.  By the inductive
hypothesis, there are parsings of $T_1, \ldots, T_k$ as p-terms.  Thus
there is a parsing of $F$ as a g-formula, and by promotion, as a
p-formula.
\end{enumerate}
\end{enumerate}

The theorem follows directly from the induction hypothesis.
\end{proof}

\subsection{Diverse Interpretations}
Let $\tset$ be a set of terms, where a term may be either a g-term or
a p-term.  We consider two terms to be distinct only if they differ
syntactically.  An expression may therefore contain multiple instances
of a single term.  We classify terms as either p-function
applications, g-function applications, or $\ITE$ terms, according to
their top-level operation. The first two categories are collectively
referred to as function application terms.  For any g-formula or
p-formula $F$, define $\tset(F)$ as the set of all function
application terms occurring in $F$.

An interpretation $\interp$ partitions a term set $\tset$ into a set of
equivalence classes, where terms $T_1$ and $T_2$ are equivalent under
$\interp$, written $T_1 \equivi T_2$ when $\einterp{T_1} =
\einterp{T_2}$.  Interpretation $\interpp$ is said to be a {\em
refinement} of $\interp$ for term set $\tset$ when $T_1 \equivip T_2 \Rightarrow T_1 \equivi T_2$ for every pair of terms $T_1$ and $T_2$ in
$\tset$.  $\interpp$ is a {\em proper} refinement of $\interp$ for
$\tset$ when it is a refinement and there is at least one pair of
terms $T_1, T_2 \in \tset$ such that $T_1 \equivi T_2$, but $T_1
\nequivip T_2$.

Let $\Sym$ denote a subset of the function symbols in p-formula $F$.  An
interpretation $\interp$ is said to be {\em diverse} for $F$ with respect to
$\Sym$ when it provides a maximal partitioning of the function
application terms in $\tset(F)$ having a top-level function symbol from
$\Sym$
relative to each other and to the other function
application terms, but subject to the constraints of functional
consistency.  That is, for $T_1$ of the form $f(T_{1,1}, \ldots,
T_{1,k})$, where $f \in \Sym$, an interpretation $\interp$ is 
diverse with respect to $\Sym$ if $\interp$
has $T_1 \equivi T_2$ only in the case where
$T_2$ is also a term of the form $f(T_{2,1}, \ldots, T_{2,k})$, and
$T_{1,i} \equivi T_{2,i}$ for all $i$ such that $1 \leq i \leq k$.  If
we let $\Symp(F)$ denote the set of all p-function symbols in $F$, then
interpretation $\interp$ is said to be {\em maximally diverse} when it
is diverse with respect to $\Symp(F)$.
Note that in a maximally diverse interpretation, the
p-function application terms for a given function symbol must
be in separate equivalence classes from those for any other p-function
or g-function symbol.

\begin{table}
\begin{center}
\begin{tabular}{|l|l|l|}
\hline
I1 & $\{x,y\}, \{g_1\} \{g_2\}, \{g_3\}, \{h_1\}, \{h_2\}$ 
  & Inconsistent \\
I2 & $\{x\}, \{y\}, \{g_1,g_2\}, \{g_3\}, \{h_1\}, \{h_2\}$ 
  & Inconsistent \\

\hline
C1 & $\{x\}, \{y\}, \{g_1,g_2\}, \{g_3\},
   \{h_1,h_2\}$ 
  & Diverse w.r.t.~$x$,$y$,$h$ \\
C2 & $\{x, g_3\}, \{y\}, \{g_1\}, \{g_2\},
  \{h_1\}, \{h_2\}$ 
  & Diverse w.r.t.~$y$, $h$\\
\hline
D1 & $\{x\}, \{y\}, \{g_1\}, \{g_2\}, \{g_3\},
  \{h_1\}, \{h_2\}$ 
  & Diverse w.r.t.~$x$, $y$, $g$, $h$\\
D2 &$\{x,y\}, \{g_1,g_2\}, \{g_3\},
   \{h_1,h_2\}$ 
  & Diverse w.r.t.~$g$, $h$\\
\hline
\end{tabular}
\end{center}
\caption{Example Partitionings of Terms 
$x$, $y$, $g_1 \doteq g(x)$, $g_2 \doteq g(y)$, $g_3 \doteq g(g(x))$,
$h_1 \doteq h(g(x),g(g(x)))$, and $h_2 \doteq h(g(y),g(g(x)))$.}
\label{partition-table}
\end{table}

As an example, consider the p-formula $\egF$ given in Equation
\ref{eg-formula}.  There are seven distinct function
application terms identified as follows: 
\begin{center}
\begin{tabular}{|c|c|c|c|c|c|c|}
\hline
$x$ & $y$ & $g_1$ & $g_2$ & $g_3$ & $h_1$ & $h_2$\\
\hline
$x$ & $y$ & $g(x)$ & $g(y)$ & $g(g(x))$ & $h(g(x),g(g(x)))$ & $h(g(y),g(g(x)))$ \\
\hline
\end{tabular}
\end{center}
Table \ref{partition-table} shows 6 of the 877 different ways to
partition seven objects into equivalence classes.  Many of these violate
functional consistency.  For example, the partitioning I1 describes a
case where $x$ and $y$ are equal, but $g(x)$ and $g(y)$ are not.
Similarly, partitioning I2 describes a case where 
$g(x)$ and $g(y)$ are equal, but
$h(g(x),g(g(x)))$ and $h(g(y),g(g(x)))$ are not.

Eliminating the inconsistent cases gives 384 partitionings.  Many of
these do not arise from maximally diverse interpretations, however.
For example, partitioning C1 arises from an interpretation that is not
diverse with respect to $g$, while partitioning C2 arises from an
interpretation that is not diverse with respect to $h$.  In fact,
there are only two partitionings: D1 and D2 that arise from maximally
diverse interpretations.  Partition D1 corresponds to an
interpretation that is diverse with respect to all of its function
symbols.  Partition D2 is diverse with respect to both $g$ and $h$,
even though terms $g_1$ and $g_2$ are in the same class, as are $h_1$
and $h_2$.  Both of these groupings are forced by functional
consistency: having $x = y$ forces $g(x) = g(y)$, which in turn forces
$h(g(x),g(g(x))) = h(g(y),g(g(x)))$.  Since $g$ and $h$ are the only
p-function symbols, D2 is maximally diverse.

The following is the central result of the paper.

\begin{theorem}
A p-formula $F$ is universally valid if and only if it is 
true in all maximally diverse interpretations.
\label{diverse-theorem}
\end{theorem}

First, it is clear that if $F$ is universally valid, $F$ is true in
all maximally diverse interpretations.  We prove via the following
two lemmas that if $F$ is true in all maximally diverse interpretations it
is universally valid.

\begin{lemma}
If interpretation $\jinterp$ is not maximally diverse for p-formula
$F$, then there is an interpretation $\jinterpp$ that is a proper
refinement of $\jinterp$ such that $\ejinterpp{F} \Rightarrow \ejinterp{F}$.
\label{refine-lemma}
\end{lemma}

\begin{proof}
Let $T_1$ be a term occurring in $F$ of the form $f_1(T_{1,1}, \ldots,
T_{1,k_1})$, where $f_1$ is a p-function symbol.  Let $T_2$ be a term
occurring in $F$ of the form $f_2(T_{2,1}, \ldots, T_{2,k_2})$, where
$f_2$ may be either a p-function or a g-function symbol.  Assume
furthermore that $\ejinterp{T_1}$ and $\ejinterp{T_2}$ both equal $z$,
but that either symbols $f_1$ and $f_2$ differ, or $\ejinterp{T_{1,i}} \not =
\ejinterp{T_{2,i}}$ for some value of $i$.

Let $z'$ be a value not in $\domain$, and define a new domain
$\domainp \doteq \domain \cup \{z'\}$.  Our strategy is to
construct an interpretation $\jinterpp$ over $\domainp$ that partitions the
terms in
$\tset(F)$ in the same way as $\jinterp$, except that it splits the class
containing terms $T_1$ and $T_2$ into two parts---one containing $T_1$
and evaluating to $z'$, and the other containing $T_2$ and evaluating
to $z$.

Define function
$\remap \colon \domainp \rightarrow \domain$ to map elements of $\domainp$ back to
their counterparts in $\domain$,
i.e., $\remap(z') = z$, while all other values of $x$ give $\remap(x)$
equal to $x$.

For p-function symbol $f_1$,
define $\ajinterpp{f_1}$ as:
\begin{eqnarray*}
\ajinterpp{f_1}(x_1, \ldots, x_{k_1}) & \doteq &
\left \{
\begin{array}{ll}
z', & \remap(x_i) = \ejinterp{T_{1,i}}, \; 1\leq i \leq k_1 \\
\ajinterp{f_1}(\remap(x_1), \ldots, \remap(x_{k_1})), & \mbox{otherwise}
\end{array}
\right .
\end{eqnarray*}

For other function and predicate symbols, $\jinterpp$ is defined to preserve
the functionality of interpretation $\jinterp$, while also treating argument
values of $z'$ the same as $z$.
That is, $\ajinterpp{f}$ for
function symbol $f$ having $\aorder{f}$ equal to 
$k$ is defined such that
$\ajinterpp{f}(x_1, \ldots, x_{k}) 
=\ajinterp{f}(\remap(x_1), \ldots, \remap(x_{k}))$.
Similarly, $\ajinterpp{p}$ for 
predicate symbol $p$ having $\aorder{p}$ equal to $k$ is defined such that
$\ajinterpp{p}(x_1, \ldots, x_{k})
= \ajinterp{p}(\remap(x_1), \ldots, \remap(x_{k}))$.

We claim the following properties for the different forms of subexpressions occurring in
$F$:
\begin{enumerate}
\item For every g-formula $G$: $\ejinterpp{G} = \ejinterp{G}$
\item For every g-term $T$: $\ejinterpp{T} = \ejinterp{T}$
\item For every p-term $T$: $\remap(\ejinterpp{T}) = \ejinterp{T}$
\item For every p-formula $G$: $\ejinterpp{G} \Rightarrow \ejinterp{G}$
\item $\ejinterpp{T_1} = z'$ and $\ejinterpp{T_2} = z$.
\end{enumerate}

Informally, interpretation $\jinterpp$ maintains the values of all
g-terms and g-formulas as occur under interpretation $\jinterp$.  It also
maintains the values of all p-terms, except those in the class
containing terms $T_1$ and $T_2$.  These p-terms are split into some
 having valuation $z$ and others
having valuation $z'$.  With respect to p-formulas, consider first an
equation of the form $\compare{S_1}{S_2}$ where $S_1$ and $S_2$ are
p-terms.  The equation will yield the same value under both
interpretations except under the condition that $S_1$ and $S_2$ are
split into different parts of the class that originally evaluated to
$z$, in which case the equation will yield $\true$ under $\jinterp$,
but $\false$ under $\jinterpp$.  Thus, although this equation can yield
different values under the two interpretations, we always have that
$\ejinterpp{\compare{S_1}{S_2}} \Rightarrow
\ejinterp{\compare{S_1}{S_2}}$.  This implication relation is
preserved by conjunctions and disjunctions of p-formulas, due to the
monotonicity of these operations.

We will now present this argument formally.  Most of the cases are
straightforward; we indicate those that are ``interesting.''
We prove hypotheses 1 to
4 above by simultaneous induction on the expression structures.

For the base cases, we have:
\begin{enumerate}
\item G-formula: $\jinterpp[\true]=\jinterp[\true]$,
   $\jinterpp[\false]=\jinterp[\false]$, and $\jinterpp[a] =\jinterp[a]$ for
any propositional variable $a$.

\item G-term: If $v$ is a g-function symbol of zero order, then
   $\ajinterpp{v} = \ajinterp{v}$. 

\item P-term: If $v$ is a p-function symbol of zero order, then by the
   definition of $\jinterpp$, $\remap(\ajinterpp{v}) = \ajinterp{v}$.

\item P-formula: same as g-formula.
\end{enumerate}

For the inductive step, we prove that hypotheses 1 through 4 hold for
an expression given that they hold for all of its subexpressions.

\begin{enumerate}
\item 
G-formula: There are several cases, depending on the form of $G$.
\begin{enumerate}
\item
Suppose $G$ has one of the forms $\neg G_1$, $G_1 \land G_2$, $G_1
\lor G_2$, where $G_1$ and $G_2$ are g-formulas.  By the inductive
hypothesis, $\jinterpp[G_1] = \jinterp[G_1]$, and $\jinterpp[G_2] =
\jinterp[G_2]$.  It follows that $\jinterpp[\neg G_1] = \jinterp[\neg
G_1]$, $\jinterpp[G_1 \land G_2] = \jinterp[G_1 \land G_2]$, and
$\jinterpp[G_1 \lor G_2] = \jinterp[G_1 \lor G_2]$.

\item
Suppose $G$ has the form $\compare{S_1}{S_2}$, where $S_1, S_2$ are
g-terms. By the inductive hypothesis on g-terms, $\jinterpp[S_1] =
\jinterp[S_1]$, and $\jinterpp[S_2] = \jinterp[S_2]$.  It follows that
$\jinterpp[\compare{S_1}{S_2}] = \jinterp[\compare{S_1}{S_2}]$.

\item 
The remaining case is that $G$ is a predicate application of the form
$p(S_1, \ldots, S_k)$, where $p$ is a predicate symbol of order $k$,
and $S_1, \ldots, S_k$, are p-terms.  By the inductive hypothesis for
p-terms, we have $\remap(\jinterpp[S_i]) = \jinterp[S_i]$, for $i = 1 \ldots k$.
By the definition of $\jinterpp$,
\[
\begin{array}{rcl}
\jinterpp[p(S_1, \ldots, S_k)] & = &
\jinterpp(p)(\jinterpp[S_1], \ldots, \jinterpp[S_k]) \\
& = & \jinterp(p)(\remap(\jinterpp[S_1]), \ldots, \remap(\jinterpp[S_k])) \\
& = & \jinterp(p)(\jinterp[S_1], \ldots, \jinterp[S_k]) \\
& = & \jinterp[p(S_1, \ldots, S_k)].
\end{array}
\]

\end{enumerate}

\item
G-term: There are two cases.

\begin{enumerate}

\item 
Suppose $T$ has the form $\ITE(G, S_1, S_2)$, where $G$ is a
g-formula, and $S_1$ and $S_2$ are g-terms.  By the inductive hypothesis, we
have $\jinterpp[G] = \jinterp[G]$, $\jinterpp[S_1] = \jinterp[S_1]$, and
$\jinterpp[S_2] = \jinterp[S_2]$.  Then $\jinterpp[\ITE(G, S_1, S_2)] =
\jinterp[\ITE(G, S_1, S_2)]$.

\item
Suppose $T$ has the form $f(S_1, \ldots, S_k)$, where $f$ is a
g-function symbol of order $k$ and $S_1, \ldots, S_k$ are p-terms.
By the inductive hypothesis, $\remap(\jinterpp[S_i]) = \jinterp[S_i]$, for $i
= 1, \ldots, k$.  Then we have,
\[
\begin{array}{rcl}
\jinterpp[f(S_1, \ldots, S_k)] & = & 
\jinterpp(f)(\jinterpp[S_1], \ldots, \jinterpp[S_k]) \\
& = & \jinterp(f)(\remap(\jinterpp[S_1]), \ldots, \remap(\jinterpp[S_k])) \\
& = & \jinterp(f)(\jinterp[S_1], \ldots, \jinterp[S_k]) \\
& = & \jinterp[f(S_1, \ldots, S_k)].
\end{array}
\]

\end{enumerate}

\item
P-term: There are three cases.

\begin{enumerate}
\item
Suppose $T$ is a g-term.  By the inductive hypothesis, $\jinterpp[T] =
\jinterp[T]$.  Since $\jinterp[T]$ cannot be equal to $z'$, it must be 
the case that $\remap(\jinterpp[T]) = \jinterp[T]$.

\item
Suppose $T$ has the form $\ITE(G, S_1, S_2)$, where $G$ is a
g-formula, and $S_1$ and $S_2$ are p-terms.  By the inductive hypothesis,
$\jinterpp[G] = \jinterp[G]$, $\remap(\jinterpp[S_1]) = \jinterp[S_1]$, and
$\remap(\jinterpp[S_2] = \jinterp[S_2])$.  It follows that
\[
\begin{array}{rcl}
\remap(\jinterpp[\ITE(G, S_1, S_2)]) & = & 
{\rm \ if\ } \jinterpp[G] {\rm \ then\ } \remap(\jinterpp[S_1]) {\rm \
else\ } \remap(\jinterpp[S_2]) \\
& = & {\rm \ if\ } \jinterp[G] {\rm \ then\ } \jinterp[S_1] 
{\rm \ else\ } \jinterp[S_2] \\
& = & \jinterp[\ITE(G, S_1, S_2)].
\end{array}
\]

\item
{\bf [Important case:]}
Suppose that $T$ has the form $f(S_1, \ldots, S_k)$, where $f$ is a
p-function symbol of order $k$ and $S_1, \ldots, S_k$ are p-terms.
Here, we have to consider two cases.  The first case is that 
the following two conditions hold: (1) $f$ is
the function symbol $f_1$, i.e., the function symbol of the term $T_1$
mentioned at the beginning of the proof of this lemma, and 
(2) $\remap(S_i) = \jinterp[T_{1,i}]$, for $1 \leq i \leq k$.  If these
two conditions hold, then by the definition of $\jinterpp$, 
$\jinterpp[f_1(S_1, \ldots, S_k)] = z'$, while $\jinterp[f_1(S_1, \ldots,
S_k)] = z$.  Since $\remap(z') = z$, we have $\remap(\jinterpp[f_1(S_1, \ldots,
S_k)]) = \jinterp[f_1(S_1, \ldots, S_k)]$.

The second case is when one of the two conditions mentioned
above does not hold.  
The proof of this case is identical to the proof of case 2(b) above.

\end{enumerate}

\item
P-formula:  There are three cases.
\begin{enumerate}

\item If the p-formula $G$ is a g-formula, then by the inductive
hypothesis, $\jinterpp[G] = \jinterp[G]$, so $\jinterpp[G] \Rightarrow
\jinterp[G]$.  

\item Suppose $G$ has one of the forms $G_1 \land G_2$, or $G_1 \lor
G_2$, where $G_1, G_2$ are p-formulas.  By the inductive hypothesis,
$\jinterpp[G_1] \Rightarrow \jinterp[G_1]$, and $\jinterpp[G_2] \Rightarrow
\jinterp[G_2]$.  Thus we have 
\[
\begin{array}{rl}
\jinterpp[G_1 \land G_2] & =  \jinterpp[G_1] \land \jinterpp[G_2] \\
& \Rightarrow  \jinterp[G_1] \land \jinterp[G_2] \\
& = \jinterp[G_1 \land G_2],
\end{array}
\]
so $\jinterpp[G_1 \land G_2] \Rightarrow \jinterp[G_1 \land G_2]$.
The proof for $G_1 \lor G_2$ is the same.

\item 
{\bf [Important case:]}
Finally, we consider the case that $G$ is a p-formula of the form $\compare{S_1}{S_2}$,
where $S_1$ and 
$S_2$, are p-terms.  By the inductive hypothesis, we have that if $\jinterpp[S_i] = z'$, then
$\jinterp[S_i] = z$, for
$i=1,2$.  Also, by the definition of $h$, we have that if
$\jinterpp[S_i]$ does not equal $z'$, then $\jinterpp[S_i] = \jinterp[S_i]$.  Now,
we consider cases depending on whether $\jinterpp[S_1]$ or $\jinterpp[S_2]$
are equal to $z'$.  If both terms are equal to $z'$ in $\jinterpp$,
then both $\jinterp[S_1]$ and $\jinterp[S_2]$ must be equal to $z$, so
the equation is true in both $\jinterpp$ and $\jinterp$.  If neither
$\jinterpp[S_1]$ nor $\jinterpp[S_2]$ is equal to $z'$, then
$\jinterpp[S_1] = \jinterp[S_1]$ and $\jinterpp[S_2] = \jinterp[S_2]$,
so the equation has the
same truth value in $\jinterpp$ and $\jinterp$.  
The last case is that exactly one of the p-terms is
equal to $z'$ in $\jinterpp$.  In this case, the equation is false in
$\jinterpp$, so we have $\jinterpp[G] \Rightarrow \jinterp[G]$.  This
completes the inductive proof.

\end{enumerate}

\end{enumerate}

Property 5 above, which implies that $\jinterpp$ is a proper
refinement, is a consequence of the definition of $\jinterpp$ and the
inductive properties 2 and 3.  First, we show that $\jinterpp[T_1] =
z'$.  By definition, $\jinterpp[T_1] = \jinterpp(f_1)(\jinterpp[T_{1,1}],
\ldots, \jinterpp[T_{1, k_1}])$.  By property 3 on
p-terms, we can assume $\remap(\jinterpp[T_{1,i}]) = \jinterp[T_{1,i}]$, for
all $i$ in the range $1 \leq i \leq k_1$.  By the definition of $\jinterpp(f_1)$,
we have $ \jinterpp(f_1)(\jinterpp[T_{1,1}], \ldots, \jinterpp[T_{1,
k_1}]) = z'$.

The proof that $\jinterpp[T_2] = z$ is in two cases, depending on
whether $T_1$ and $T_2$ are applications of the same function symbol.

\begin{enumerate}
\item 
First, consider the case that $T_1 = f_1(T_{1,1}, \ldots, T_{1,k_1})$
and $T_2 = f_2(T_{2,1}, \ldots, T_{2,k_2})$, where $f_1$ and $f_2$ are
different function symbols.  In this case,
\[
\begin{array}{rcl}
\jinterpp[T_2] & = & \jinterpp(f_2)(\jinterpp[T_{2,1}], \ldots,
\jinterpp[T_{2,k_2}])   \\
& = & \jinterp(f_2)(\remap(\jinterpp[T_{2,1}]), \ldots,
\remap(\jinterpp[T_{2,k_2}])),
 \mbox{by the definition of $\jinterpp(f_2)$}   \\
& = & \jinterp(f_2)(\jinterp[T_{2,1}], \ldots, \jinterp[T_{2,k_2}]),
 \mbox{by the inductive hypothesis}   \\
& = & \jinterp[f_2(T_{2,1}, \ldots, T_{2,k_2})]  \\
& = & z.
\end{array}
\]

\item
Finally, we have the case that $f_1$ and $f_2$ are the same function
symbol, and there is some value of $l$ with $1 \leq l \leq k_1$, such that
$\jinterp[T_{1,l}]$ does not equal $\jinterp[T_{2,l}]$.  Here, we have:
\[
\jinterpp[f_1(T_{2,1}, \ldots, T_{2,k_2})] =
\jinterpp(f_1)(\jinterpp[T_{2,1}], \ldots, \jinterpp[T_{2,k_2}])
\]
By property 3, $\remap(\jinterpp[T_{2,i}]) =
\jinterp[T_{2,i}]$, for all $i$ such that $1 \leq i \leq k_1$.  Since
$\jinterp[T_{1,l}]$ does not equal $\jinterp[T_{2,l}]$, the value of the above
application of $\jinterpp(f_1)$ is:
\[
\begin{array}{rcl}
\jinterpp(f_1)(\jinterpp[T_{2,1}], \ldots, \jinterpp[T_{2,k_2}]) & = &
\jinterp(f_1)( \remap(\jinterpp[T_{2,1}]), \ldots, \remap(\jinterpp[T_{2,k_2}]))  \\
& = & \jinterp(f_1)(\jinterp[T_{2,1}], \ldots, \jinterp[T_{2,k_2}])  \\
& = & \jinterp[f_1(T_{2,1}, \ldots, T_{2,k_2})]  \\
& = & z
\end{array}
\]
\end{enumerate}
\end{proof}

\begin{lemma}
For any interpretation $\interp$ and p-formula $F$, there is a
maximally diverse interpretation $\interps$ for $F$ such that
$\einterps{F} \Rightarrow \einterp{F}$.
\label{diverse-lemma}
\end{lemma}

\begin{proof}
Starting with interpretation $\interp_0$ equal to $\interp$, we define
a sequence of interpretations $\interp_0, \interp_1, \ldots$ by
repeatedly applying the construction of Lemma \ref{refine-lemma}.
That is, we derive each interpretation $\interp_{i+1}$ from its
predecessor $\interp_{i}$ by letting $\jinterp = \interp_{i}$ and
letting $\interp_{i+1} = \jinterpp$.  Interpretation $\interp_{i+1}$
is a proper refinement of its predecessor $\interp_i$ such that
$\eval{\interp_{i+1}}{F} \Rightarrow \eval{\interp_{i}}{F}$.  At some
step $n$, we must reach a maximally diverse interpretation
$\interp_n$, because our set $\tset(F)$ is finite and therefore can
be properly refined only a finite number of times.  We then let
$\interps$ be $\interp_n$.  We can see that $\einterps{F} =
\eval{\interp_{n}}{F} \Rightarrow \cdots \Rightarrow
\eval{\interp_{0}}{F} = \einterp{F}$, and hence $\einterps{F}
\Rightarrow \einterp{F}$.
\end{proof}

The completion of the proof of Theorem \ref{diverse-theorem} follows
directly from Lemma \ref{diverse-lemma}.  That is, if we start with
any interpretation $\interp$ for p-formula $F$, we can construct a maximally
diverse interpretation $\interps$ such that $\einterps{F} \Rightarrow
\einterp{F}$.  Assuming $F$ is true under all maximally diverse
interpretations, $\einterps{F}$ must hold, and since
$\einterps{F} \Rightarrow \einterp{F}$, $\einterp{F}$ must hold as well.

\subsection{Exploiting Positive Equality in a Decision Procedure}

A decision procedure for PEUF must determine whether a given p-formula
is universally valid.  The procedure can significantly reduce the
range of possible interpretations it must consider by exploiting the
maximal diversity property.
Theorem \ref{diverse-theorem} shows that we can consider only
interpretations in which the values produced by the application
of any p-function symbol differ from those produced by the
applications of any other p-function or g-function symbol.
\comment{
Thus, if
there are $m$ different p-function symbols $f_1, \ldots, f_m$, we can
partition domain $\domain$ into subsets $\domain_0, \domain_1, \ldots,
\domain_m$, where applications of g-function symbols yield only values
in $\domain_0$, and applications of each p-function symbol $f_i$ yields only
values in $\domain_i$.  Furthermore, we can consider each application
of p-function symbol $f_i$ to yield a distinct value, except when its
arguments match those of some other application.
}
We can therefore consider the different p-function symbols to yield
values over domains disjoint with one another and with the domain of
g-function values.  In addition, we can consider each application of a
p-function symbol to yield a distinct value, except when its arguments
match those of some other application.

\section{Eliminating Function Applications}

Most work on transforming EUF into propositional logic has used the
method described by Ackermann to eliminate applications of functions
of nonzero order \cite{ackermann-54}.  In this scheme, each function
application term is replaced by a new domain variable and constraints
are added to the formula expressing functional consistency.  Our
approach also introduces new domain variables, but it replaces each
function application term with a nested $\ITE$ structure that directly
captures the effects of functional consistency.  As we will show, our
approach can readily exploit the maximal diversity property, while
Ackermann's cannot.

In the presentation of our method for eliminating function and
predicate applications, we initially consider formulas in EUF\@.  We
then show how our elimination method can exploit maximal diversity in
PEUF formulas.

\subsection{Function Application Elimination Example}

\label{example-section}

\begin{center}
\begin{figure}
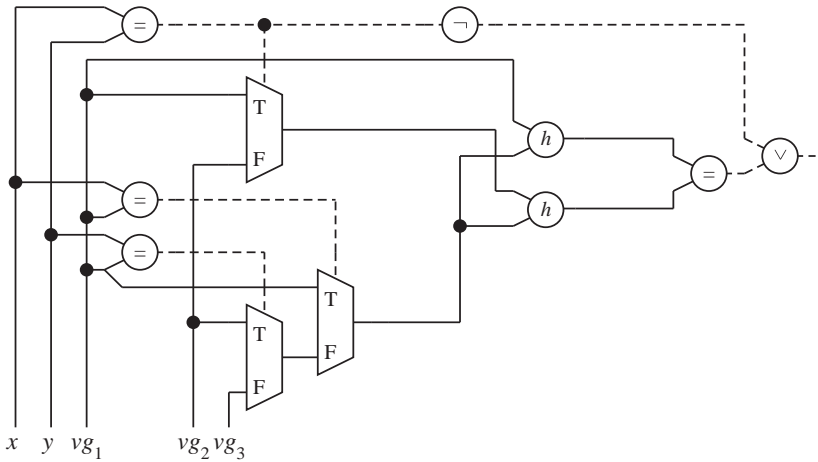
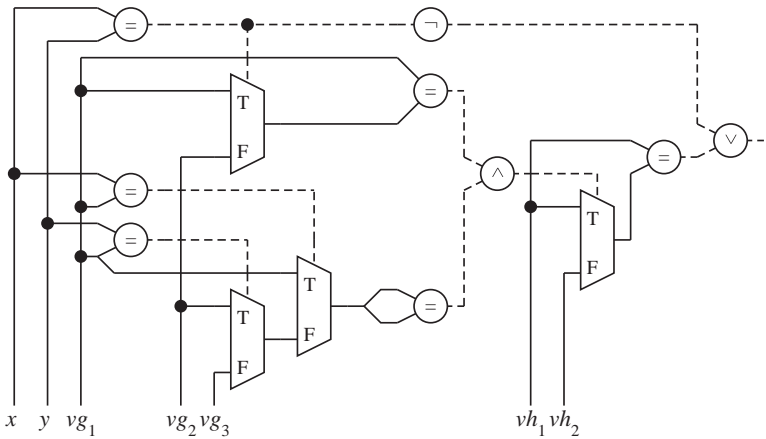

Initial formula:\\[1.5ex]
\centerline{\psfig{figure=\findfigure{eg-original},width=4.29in}}
After removing applications of function symbol $g$:\\[1.5ex]
\centerline{\psfig{figure=\findfigure{eg-partial},width=4.29in}}
After removing applications of function symbol $h$:\\[1.5ex]
\centerline{\psfig{figure=\findfigure{eg-final},width=4.29in}}
\caption{Removing Function Applications from $\egF$.}
\label{ite-expansion-figure}
\end{figure}
\end{center}

We demonstrate our technique for replacing function applications by
domain variables using formula $\egF$ (Equation \ref{eg-formula}) as
an example, as illustrated in Figure \ref{ite-expansion-figure}.
First consider the three applications of function symbol $g$: $g(x)$,
$g(y)$, and $g(g(x))$, which we identify as terms $T_1$, $T_2$, and
$T_3$, respectively.  Let $\vg_1$, $\vg_2$, and $\vg_3$ be new domain
variables.  We generate new terms $U_1$, $U_2$, and $U_3$ as follows:
\begin{eqnarray}
U_1 & \doteq & \vg_1 
\label{variable-transform-equation} \\
U_2 & \doteq & \ITE(\compare{y}{x}, \vg_1, \vg_2) \nonumber\\
U_3 & \doteq & \ITE(\compare{vg_1}{x}, \vg_1, \ITE(\compare{vg_1}{y},
\vg_2, \vg_3)) \nonumber
\end{eqnarray}
We use variable $vg_1$, the translation of $g(x)$, to
represent the argument to the outer application of function symbol $g$
in the term $g(g(x))$.  In general, we must always process nested
applications of a given function symbol working from the innermost to
the outermost.  Given terms $U_1$, $U_2$, and $U_3$, we eliminate
the function applications by replacing each instance of $T_i$ in the
formula by $U_i$ for $1 \leq i \leq 3$, as shown in the middle part of
Figure \ref{ite-expansion-figure}.  We use multiplexors in our
schematic diagrams to represent $\ITE$ operations.

\begin{table}
\begin{center}
\begin{tabular}{|c|ccc|}
\hline
$\equivip$ & $\einterpp{U_1}$ & $\einterpp{U_2}$ & $\einterpp{U_3}$ \\
\hline
$\{ x \}, \{ y \}, \{ g(x) \}$ &   $1$ & $2$ & $3$ \\
$\{ x , y \}, \{ g(x) \}$ &   $1$ & $1$ & $3$ \\
$\{ x \}, \{ y , g(x) \}$ &   $1$ & $2$ & $2$ \\
$\{ x , g(x) \}, \{ y \}$ &   $1$ & $2$ & $1$ \\
$\{ x ,  y ,  g(x) \}$ &   $1$ & $1$ & $1$ \\
\hline
\end{tabular}
\end{center}
\caption{Possible valuations of terms in Equation \protect{\ref{variable-transform-equation} when each variable $\vg_i$ is assigned value $i$}.}
\label{transform-example-table}
\end{table}

Observe that as we consider interpretations with different values for
 variables $\vg_1$, $\vg_2$, and $\vg_3$ in Equation
 \ref{variable-transform-equation}, we implicitly cover all values
 that an interpretation of function symbol $g$ in formula $\egF$ may
 yield for the three arguments.  The nested $\ITE$ structure shown in
 Equation \ref{variable-transform-equation} enforces functional
 consistency.  For example, consider an arbitrary interpretation
 $\interp$ of the symbols in $\egF$.  Define interpretation $\interpp$
 to be identical to $\interp$ for the symbols in $\egF$ and in
 addition to assign values $1$, $2$, and $3$ to domain variables
 $\vg_1$, $\vg_2$, and $\vg_3$, respectively.  Table
 \ref{transform-example-table} shows the possible valuations of the
 three terms of Equation \ref{variable-transform-equation} under
 $\interpp$.  For each possible partitioning by $\interps$ of
 arguments $x$, $y$, and $g(x)$ into equivalence classes, we get
 $\einterpp{U_i} = \einterpp{U_j}$ if an only if the arguments to
 function application terms $T_i$ and $T_j$ are equal under $\interp$.

We remove the two applications of function symbol $h$ by a similar
process.  That is, we introduce two new domain variables $vh_1$ and
$vh_2$.  We replace the first application of $h$ by $\vh_1$ and the
second by an $\ITE$ term that compares the arguments of the two
function applications, yielding $vh_1$ if they are equal and $\vh_2$
if they are not.  The final form is illustrated in the bottom part of
Figure \ref{ite-expansion-figure}.  The translation of predicate
applications is similar, introducing a new propositional variable for
each application.  After removing all applications of function and
predicate symbols of nonzero order, we are left with a formula $\segF$
containing only domain and propositional variables.

\subsection{Algorithm for Eliminating Function and Predicate Applications}
\label{ite-expansion-section}

The general translation procedure follows the form shown for our
example.  It iterates through the function and predicate symbols of
nonzero order.  On each iteration it eliminates all occurrences of a
given symbol.  At the end we are left with a formula containing only
domain and propositional variables.

\begin{figure}
Initial p-formula showing $f$-order contours:\\[1.5ex]
\centerline{\psfig{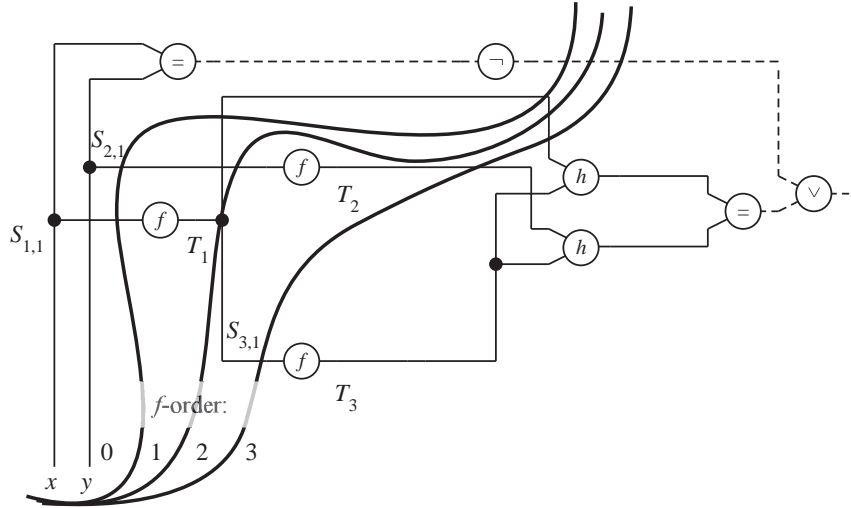}}
After removing applications of function symbol $f$:\\[1.5ex]
\centerline{\psfig{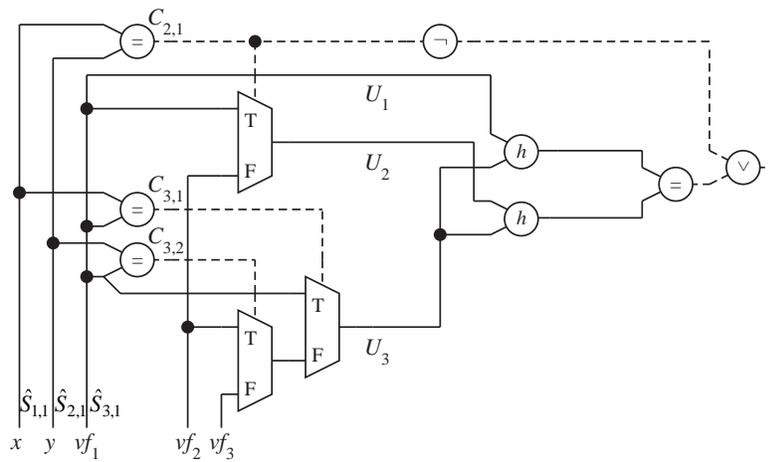}}
\caption{Illustration of Function Application Removal}
\label{eg-forder-figure}
\end{figure}

The following is a detailed description of the process required to
eliminate all instances of a single function symbol $f$ having order
$k > 0$ from a formula $G$.  We use the variant of formula $\egF$ shown
schematically at the top of Figure \ref{eg-forder-figure}.
In this variant, we have replaced function
symbol $g$ with $f$.
In the sequel, if $E$ is an expression and
$T$ and $U$ are terms, we will write $E[T \leftarrow U]$ for the
result of substituting $U$ for each instance of $T$ in $E$.  Let
$T_{1}, \ldots, T_{n}$ denote the syntactically distinct terms
occurring in formula $G$ having the application of $f$ as the top
level operation.  We refer to these as ``$f$-application'' terms.  Let
the arguments to $f$ in $f$-application term $T_i$ be the terms
$S_{i,1}, \ldots, S_{i,k}$, so that $T_i$ has the form $f(S_{i,1},
\ldots, S_{i,k})$.  Assume the terms $T_1, \ldots, T_n$ are ordered
such that if $T_{i}$ occurs as a subexpression of $T_{j}$ then $i <
j$. In our example the $f$-application terms are: $T_1 \doteq f(x)$,
$T_2 \doteq f(y)$ and $T_3 \doteq f(f(x))$.  These terms have
arguments: $S_{1,1} \doteq x$, $S_{2,1} \doteq y$, and $S_{3,1} \doteq
f(x)$.

The translation processes the $f$-application terms in order,
such that on step $i$ it replaces all occurrences of the $\ordth{i}$
application of function symbol $f$ by a nested $\ITE$ term.  Let
$\vf_1, \ldots, \vf_n$ be a new set of domain variables not
occurring in $F$.  We use these to encode
the possible values returned by the $f$-application terms.

For any subexpression $E$ in $G$ define its integer-valued $f$-order,
denoted $\forder{E}$, as the highest index $i$ of an $f$-application
term $T_i$ occurring in $E$.  If no $f$-application terms occur in
$E$, its $f$-order is defined to be 0.  By our ordering of the
$f$-application terms, any argument $S_{i,l}$ to $f$-application term
$T_i$ must have $\forder{S_{i,l}} < \forder{T_i}$, and therefore
$\forder{T_i} = i$.  For example, the contour lines shown in Figure
\ref{eg-forder-figure} partition the operators according to their
$f$-order values.

The transformations performed in replacing applications of function
symbol $f$ can be expressed by defining the following recurrence for
any subexpression $E$ of $G$:
\begin{equation}
\begin{array}{rcll}
\expru{0} & \doteq & E \\
\expru{i} & \doteq & \expru{i-1}[\termud{i-1}{i} \leftarrow U_i], & 1 \leq i \leq n\\
\hE & \doteq & \expru{m}, & \mbox{where}\; m = \forder{E}\\
\end{array}
\label{step-equation}
\end{equation}
In this equation, term $\termud{i-1}{i}$ is the form of the
$\ordth{i}$ $f$-application term $T_i$ after all but the topmost
application of $f$ have been eliminated.  Term $U_i$ is a nested $\ITE$
structure encoding the possible values returned by $T_i$ while
enforcing its consistency with earlier applications.  $U_i$ does not
contain any applications of function symbol $f$.  For a subexpression
$E$ with $\forder{E} = m$, its form $\expru{m}$ will contain no
applications of function symbol $f$.  We denote this form as $\hE$.
Observe that for any $i > \forder{E}$, term $\termud{i-1}{i}$ does not
occur in $\expru{i}$, and hence $\expru{i} = \hE$ for all $i \geq
\forder{E}$.  Observe also that for $f$-application term $T_i$, we have
$\hT_i = \termud{i}{i} = U_i$.

$U_i$ is defined in terms of a recursively-defined term $V_{i,j}$ as
follows:
\begin{equation}
\begin{array}{rcll}
V_{i,i} & \doteq & \vf_i, & 1 \leq i \leq n\\
V_{i,j} & \doteq & \ITE(C_{i,j}, \vf_j, V_{i,j+1}), & 1 \leq j < i \leq n\\
U_{i} & \doteq & V_{i,1}, & 1 \leq i \leq n \\
\end{array}
\label{term-equation}
\end{equation}
where for each $j < i$, formula $C_{i,j}$ 
is true iff the (transformed) arguments to the top-level
application of $f$ in the terms $T_i$ and $T_j$ have the same values:
\begin{equation}
C_{i,j} \;\;\;\doteq\;\;\; \bigwedge_{1 \leq l \leq k} \compare{\hS_{i,l}}{\hS_{j,l}}
\label{consistency-equation}
\end{equation}
Observe that the recurrence of Equation \ref{term-equation} is
well-defined, since for all argument terms of the form $S_{j,l}$ for
$1 \leq j \leq i$ and $1 \leq l \leq k$, we have $\forder{S_{j,l}} <
i$, and hence terms of the form $\hS_{j,l}$ and $\hS_{i,l}$, as well
as term $V_{i,j+1}$ are available when we define $V_{i,j}$.

The lower part of Figure \ref{eg-forder-figure} shows the result of
removing the three applications of $f$ from our example formula.
First, we have $U_1 \doteq \vf_1$, giving translated function arguments:
$\hS_{1,1} \doteq x$, $\hS_{2,1} \doteq y$, and $\hS_{3,1} \doteq \vf_1$.
The comparison formulas are then:
$C_{2,1} \doteq (\compare{y}{x})$,
$C_{3,1} \doteq (\compare{\vf_1}{x})$,
and
$C_{3,2} \doteq (\compare{\vf_1}{y})$.  From these we get translated terms:
\begin{eqnarray*}
U_2 & \doteq & \ITE(\compare{y}{x}, \vf_1, \vf_2)\\
U_3 & \doteq & \ITE(\compare{\vf_1}{x}, \vf_1, \ITE(\compare{\vf_1}{y},
\vf_2, \vf_3)) 
\end{eqnarray*}

We can see that formula $\hG
\doteq \gformu{n}$ will no longer contain any applications of function
symbol $f$.  We will show that $\hG$ is universally valid if
and only if $G$ is.

In the following correctness proofs, we will use a fundamental
principle relating syntactic substitution and expression evaluation:
\begin{proposition}
For any expression $E$, pair of terms $T$, $U$, and interpretation
$\interp$ of all of the symbols in $E$, $T$, and $U$, if $\einterp{T}
= \einterp{U}$ then $\einterp{E[T \leftarrow U]} = \einterp{E}$.
\label{replace-proposition} 
\end{proposition}

We will also use the following characterization of Equation
\ref{term-equation}.  For value $i$ such that $1
\leq i \leq n$ and for interpretation $\interp$ of the symbols in
$U_i$, we define the {\em least matching value} of $i$
under interpretation $\interp$, denoted
$\lma{\interp}{i}$, as the minimum value $j$ in the range $1 \leq j
\leq i$ such that $\einterp{\hS_{j,l}} = \einterp{\hS_{i,l}}$ for all
$l$ in the range $1 \leq l \leq k$.  Observe that this value is well
defined, since $i$ forms a feasible value for $j$ in
any case.  
\begin{lemma}
For any interpretation $\interp$, $\einterp{U_i} = \ainterp{\vf_j}$,
where $j = \lma{\interp}{i}$.
\label{U-lemma}
\end{lemma}
\begin{proof}
For value $m$ in the range $1 \leq m \leq i$ define
$\lmr{\interp}{m}{i}$ as the minimum value of $j$ in the range $m \leq
j \leq i$ such that $\einterp{\hS_{j,l}} = \einterp{\hS_{i,l}}$ for
all $l$ in the range $1 \leq l \leq k$.  By this definition
$\lma{\interp}{i} = \lmr{\interp}{1}{i}$.  Observe also that if $j =
\lmr{\interp}{m}{i}$ then $\einterp{C_{i,j}} = \true$.  In addition,
for any value $m'$ in the range $m \leq m' \leq i$, if
$\lmr{\interp}{m}{i} \geq m'$, then $\lmr{\interp}{m}{i} =
\lmr{\interp}{m'}{i}$.

We prove by induction on $m$ that $\einterp{V_{i,m}} = \ainterp{\vf_j}$,
where $j = \lmr{\interp}{m}{i}$.  The base case of $m= i$ is trivial,
since $\lmr{\interp}{i}{i} = i$, and $V_{i,i} = \vf_i$.

Assuming the property holds for $m+1$, we consider two possibilities.
First, if $\lmr{\interp}{m}{i} = m$, we have $\einterp{C_{i,m}} =
\true$, and hence the top-level $\ITE$ operation in $V_{i,m}$
(Equation \ref{term-equation}) will select its first term argument
$\vf_m$, giving $\einterp{V_{i,m}} = \ainterp{\vf_m}$.  On the other hand,
if $\lmr{\interp}{m}{i} > m$, we must have $\einterp{C_{i,m}} =
\false$, and hence the top-level $\ITE$ operation in $V_{i,m}$ will
select its second term argument $V_{i,m+1}$, giving $\einterp{V_{i,m}} =
\einterp{V_{i,m+1}}$, which by the inductive hypothesis equals
$\ainterp{\vf_j}$ for $j = \lmr{\interp}{m+1}{i}$.  Since
$\lmr{\interp}{m}{i} \geq m+1$, we must also have $\lmr{\interp}{m}{i}
= \lmr{\interp}{m+1}{i}$, and hence $\einterp{V_{i,m}} =
\ainterp{\vf_j}$, where $j = \lmr{\interp}{m}{i}$.

Since $U_i$ is defined as $V_{i,1}$, our induction argument proves
that $\einterp{U_i} = \ainterp{\vf_j}$ for $j = \lmr{\interp}{1}{i} =
\lma{\interp}{i}$.
\end{proof}

\begin{lemma}
Any interpretation $\jinterp$ of the symbols in $G$ can be extended to
an interpretation $\jinterph$ of the symbols in both $G$ and $\hG$ such
that for every
subexpression $E$ of $G$, $\ejinterph{\hE} = \ejinterph{E} = \ejinterp{E}$.
\label{G-hG-lemma}
\end{lemma}

\begin{proof}
We provide a somewhat more general construction of $\jinterph$ than is
required for the proof of this lemma in anticipation of using this
construction in the proof of Lemma \ref{preserve-diverse-lemma}.
Given $\jinterp$ defined over domain $\domain$, we define $\jinterph$
over a domain $\domainh$ such that $\domainh \supseteq \domain$.

We define $\jinterph$ for the function and predicate symbols occurring
in $G$ based on their definitions in $\jinterp$.  For any function
symbol $f$ in $G$ having $\aorder{f} = k$, and any argument values
$x_1, \ldots, x_k \in \domain$, we define $\ajinterph{f}(x_1, \ldots,
x_k) \doteq \ajinterp{f}(x_1, \ldots, x_k)$.  For argument values
$x_1, \ldots, x_k \in \domainh$ such that for some $i$, $x_i \not \in
\domain$, we let $\ajinterph{f}(x_1, \ldots, x_k)$ be an arbitrary
domain value.  Similarly, for predicate symbol $p$, we define
$\ajinterph{p}$ to yield the same value as $\ajinterp{p}$ for
arguments in $\domain$ and to yield an arbitrary truth value when at
least one argument is not in $\domain$.

One can readily see that $\ejinterph{E} = \ejinterp{E}$
for every subexpression $E$ of $G$.  This takes care of the second
equality in the statement of the lemma, and hence we can concentrate
on the relation between $\ejinterph{\hE}$ and $\ejinterph{E}$ for the
remainder of the proof.

Recall that $\vf_1, \ldots, \vf_n$ are the domain variables introduced
when generating the nested $\ITE$ terms $U_1, \ldots U_n$.  Our strategy is to
define interpretations of these variables such that each $U_i$ mimics the
behavior of the original $f$-application term $T_i$ in $G$.

We consider two cases.  For the case where $\lma{\jinterph}{i} = i$,
we define $\ajinterph{\vf_i} = \ejinterph{T_{i}}$, i.e., the value of
the $\ordth{i}$ $f$-application term in $G$ under $\jinterp$.
Otherwise, we let $\ajinterph{\vf_i}$ be an arbitrary domain
value---we will show that its value does not affect the valuation of
any expression $\hE$ in $\hG$ having a counterpart $E$ in $G$.

We argue by induction on $i$ that $\ejinterph{\expru{i}} = \ejinterph{E}$
for any subexpression $E$ of $G$.  For the case where $\forder{E} \leq
i$, this hypothesis implies that $\ejinterph{\hE} = \ejinterph{E}$.  The
base case of $i=0$ is trivial, since $\expru{0}$ is defined to be $E$.

Suppose that for every $j$ in the range $1 \leq j < i$ and every subexpression $D$ of $G$, we
have $\ejinterph{\dexpru{j}} = \ejinterph{D}$, and consequently that
$\ejinterph{\hD} = \ejinterph{D}$ for the case where $\forder{D} <
i$.  We must show that for every subexpression $E$ of $G$, we have
$\ejinterph{\expru{i}} = \ejinterph{E}$.

We first focus our attention on term $T_i$ in $G$ and its counterpart
$U_i$ in $\hG$, showing that $\ejinterph{U_i} = \ejinterph{T_i}$.  The
$f$-application terms for all $j$ such that $j< i$ have $\forder{T_j}
= j < i$, and hence we can assume that $\ejinterph{U_j} =
\ejinterph{T_j}$ for these values of $j$.  Furthermore, any argument
$S_{j,l}$ to an $f$-application term for $j \leq i$ and $1 \leq l \leq
k$ has $\forder{S_{j,l}} < j \leq i$, and hence we can assume
$\ejinterph{\hS_{j,l}} = \ejinterph{S_{j,l}}$.

We consider two cases: $\lma{\jinterph}{i} = i$, and
$\lma{\jinterph}{i} < i$.  In the former case,
we have by Lemma \ref{U-lemma} that $\ejinterph{U_{i}} =
\ajinterph{\vf_i}$.  Our definition of $\ajinterph{\vf_i}$ gives
$\ejinterph{U_i} = \ajinterph{\vf_i} = \ejinterph{T_i}$.
Otherwise, suppose that $\lma{\jinterph}{i} = j < i$.  Lemma
\ref{U-lemma} shows that $\ejinterph{U_i} = \ajinterph{\vf_j}$.  We
can see that $\lma{\jinterph}{j} = j$, and hence $\ajinterph{\vf_j}$
is defined to be $\ejinterph{T_j}$.  By the definition of $\lm$ we
have $\ejinterph{\hS_{j,l}} = \ejinterph{\hS_{i,l}}$ for $1 \leq l
\leq k$.  By the induction hypothesis we have $\ejinterph{\hS_{j,l}} =
\ejinterph{S_{j,l}}$, since $\forder{S_{j,l}} < i$, and similarly that
$\ejinterph{\hS_{i,l}} = \ejinterph{S_{i,l}}$.  By transitivity we
have $\ejinterph{S_{j,l}} = \ejinterph{S_{i,l}}$ for all $l$ such that
$1 \leq l \leq k$, i.e., the arguments to $f$-application terms $T_j$
and $T_i$ have equal valuations under $\jinterp$.  Function
consistency requires that $\ejinterph{T_j} = \ejinterph{T_i}$.  From
this we can conclude that $\ejinterph{U_i} = \ejinterph{U_j} =
\ejinterph{T_j} = \ejinterph{T_i}$.  Combining these cases gives
$\ejinterph{U_i} = \ejinterph{T_i}$.

For any subexpression $E$ its form $\expru{i}$ differs from
$\expru{i-1}$ only in that all instances of term $\termud{i-1}{i}$
have been replaced by $U_i$.  We have just argued that $\ejinterph{U_i}
= \ejinterph{T_i}$, and by the induction hypothesis we have that
$\ejinterph{\termud{i-1}{i}} = \ejinterph{T_i}$, giving by transitivity
that $\ejinterph{\termud{i-1}{i}} = \ejinterph{U_i}$.  Proposition
\ref{replace-proposition} implies that $\ejinterph{\expru{i}} =
\ejinterph{\expru{i-1}}$, and our induction hypothesis gives
$\ejinterph{\expru{i-1}}= \ejinterph{E}$.  By transitivity we have
$\ejinterph{\expru{i}} = \ejinterph{E}$.

To complete the proof, we observe that our induction argument implies
that for any subexpression $E$ of $G$, $\ejinterph{\expru{m}} =
\ejinterph{E}$, including for the case where $m = \forder{E}$, giving
$\ejinterph{\hE} = \ejinterph{\expru{m}} = \ejinterph{E}$.
\end{proof}

\begin{lemma}
Any interpretation $\jinterph$ of the symbols
in $\hG$ can be extended to an interpretation $\jinterp$ of the symbols in both
$\hG$ and $G$
such that for every
subexpression $E$ of $G$,
$\ejinterp{E} = \ejinterp{\hE} = \ejinterph{\hE}$ .
\label{hG-G-lemma}
\end{lemma}

\begin{proof}
We define $\jinterp$ to be identical to $\jinterph$ for any symbol
occurring in $\hG$.  This implies that $\ejinterp{\hE} =
\ejinterph{\hE}$ for every subexpression $E$ of $G$.
This takes care of the second
equality in the statement of the lemma, and hence we can concentrate
on the relation between $\ejinterp{E}$ and $\ejinterp{\hE}$ for the
remainder of the proof.

For function symbol $f$, we define $\ajinterp{f}(x_1, \ldots, x_k)$ for
domain elements $x_1, \ldots, x_k$ as follows.  Suppose there is some
value $j$ such that $x_l = \ejinterp{\hS_{j,l}}$ for all $l$ such that
$1 \leq l \leq k$, and such that $j = \lma{\jinterph}{j}$.  Then we
define $\ajinterp{f}(x_1, \ldots, x_k)$ to be $\ajinterp{\vf_j}$.  If no
such value of $j$ exists, we let $\ajinterp{f}(x_1, \ldots, x_k)$ be
some arbitrary domain value.

We argue by induction on $i$ that $\ejinterp{E} = \ejinterp{\expru{i}}$
for any subexpression $E$ of $G$.  For the case where $\forder{E} \leq
i$, this hypothesis implies that $\ejinterp{E} = \ejinterp{\hE}$.  The
base case of $i=0$ is trivial, since $\expru{0}$ is defined to be $E$.

Suppose that for every $j$ in the range $1 \leq j < i$ and every
subexpression $D$ of $G$, we have $\ejinterp{D} =
\ejinterp{\dexpru{i}}$, and consequently that $\ejinterp{D} =
\ejinterp{\hD}$ for the case where $\forder{D} < i$.  We must show
that for every subexpression $E$ of $G$, we have $\ejinterp{E} =
\ejinterp{\expru{i}}$.

We focus initially on term $T_i$ in $G$ and its counterpart $U_i$ in $\hG$, showing that
$\ejinterp{T_i} = \ejinterp{U_i}$.  Any $f$-application
term $T_j$ for $j< i$ has $\forder{T_j} = j < i$, and hence we can
assume that $\ejinterp{T_j} = \ejinterp{\hT_j}$.  Furthermore, any
argument $S_{j,l}$ to an $f$-application term for $j \leq i$ and $1
\leq l \leq k$ has $\forder{S_{j,l}} < j \leq i$, and hence we can
assume that $\ejinterp{S_{j,l}} = \ejinterp{\hS_{j,l}}$.

We consider two cases: $\lma{\jinterph}{i} = i$, and
$\lma{\jinterph}{i} < i$.  In the former case,
we have by
Lemma \ref{U-lemma} 
that $\ejinterp{U_i} = \ajinterp{\vf_i}$.  In addition, $\ajinterp{f}$
is defined such that $\ejinterp{T_i} =
\ajinterp{f}(\ejinterp{S_{i,1}}, \ldots, \ejinterp{S_{i,k}}) =
\ajinterp{f}(\ejinterp{\hS_{i,1}}, \ldots, \ejinterp{\hS_{i,k}}) =
\ajinterp{\vf_i}$, giving $\ejinterp{T_i} = \ajinterp{\vf_i} =
\ejinterp{U_i}$.
Otherwise, suppose that $\lma{\jinterp}{i} = j < i$.  Lemma
\ref{U-lemma} shows that $\ejinterp{U_i} = \ajinterp{\vf_j}$.  We can
see that $\lma{\jinterph}{j} = j$, and hence $\ajinterp{f}$ is defined
such that $\ajinterp{f}(\ejinterp{\hS_{j,1}}, \ldots,
\ejinterp{\hS_{j,k}}) = \ajinterp{\vf_j}$.  For any $l$ such that $1
\leq l \leq k$, we also have by the definition of $\lm$ that
$\ejinterp{\hS_{j,l}} = \ejinterp{\hS_{i,l}}$.  By the induction
hypothesis we have $\ejinterp{S_{j,l}} = \ejinterp{\hS_{j,l}}$, since
$\forder{S_{j,l}} < i$, and similarly that $\ejinterp{S_{i,l}} =
\ejinterp{\hS_{i,l}}$.  By transitivity we have $\ejinterp{S_{j,l}} =
\ejinterp{S_{i,l}}$, i.e., the arguments to $f$-application terms
$T_j$ and $T_i$ have equal valuations under $\jinterp$.  Functional
consistency requires that $\ejinterp{T_j} = \ejinterp{T_i}$.  Putting
this together gives $\ejinterp{T_i} = \ejinterp{T_j} =
\ajinterp{f}(\ejinterp{S_{j,1}}, \ldots, \ejinterp{S_{j,k}}) =
\ajinterp{f}(\ejinterp{\hS_{j,1}}, \ldots, \ejinterp{\hS_{j,k}}) =
\ajinterp{\vf_j} = \ejinterp{U_i}$.

For any subexpression $E$ its form $\expru{i}$ differs from
$\expru{i-1}$ only in that all instances of term $\termud{i-1}{i}$
have been replaced by $U_i$.  We have just argued that $\ejinterp{T_i}
= \ejinterp{U_i}$, and by the induction hypothesis we have that
$\ejinterp{T_i} = \ejinterp{\termud{i-1}{i}}$, giving by transitivity
that $\ejinterp{\termud{i-1}{i}} = \ejinterp{U_i}$.  Proposition
\ref{replace-proposition} implies that $\ejinterp{\expru{i-1}} =
\ejinterp{\expru{i}}$, and our induction hypothesis gives $\ejinterp{E}
= \ejinterp{\expru{i-1}}$.  By transitivity we have $\ejinterp{E} =
\ejinterp{\expru{i}}$.

To complete the proof, we observe that our induction argument implies
that for any subexpression $E$ of $G$, $\ejinterp{E} =
\ejinterp{\expru{m}}$, including for the case where $m = \forder{E}$,
giving $\ejinterp{E} = \ejinterp{\expru{m}} = \ejinterp{\hE}$.
\end{proof}

An application of a predicate symbol having nonzero order can be
removed by a similar process, using newly generated propositional
variables to encode the possible values returned by the predicate
applications.  By an argument similar to that made in Lemma
\ref{G-hG-lemma}, we can extend an interpretation to include
interpretations of the propositional variables such that the original
and the transformed formulas have identical valuations.  Conversely,
by an argument similar to that made in Lemma \ref{hG-G-lemma}, we can
extend an interpretation to include an interpretation of the original
predicate symbol such that the original and the transformed formulas
have identical valuations.

Suppose formula $F$ contains applications $m$ different function and
predicate symbols of nonzero order.  Starting with $F_{0} \doteq F$,
we can generate a sequence of formulas $F_0, F_1, \ldots, F_m$. Each
formula $F_i$ is generated from its predecessor $F_{i-1}$ by
letting $G = F_i$ and $F_{i+1} = \hG$ in
our technique to eliminate all instances of the $\ordth{i}$
function or predicate symbol.
Let $\sF \doteq F_m$ denote the formula
that will result once we have eliminated all applications of function
and predicate symbols having nonzero order.

\begin{theorem}
For EUF formula $F$, the transformation process described above yields
a formula $\sF$ such that $F$ is universally valid if and only if
$\sF$ is universally valid.
\label{euf-transform-theorem}
\end{theorem}

\begin{proof}
{\it If:}  Assume $\sF$ is universally valid, and consider any interpretation $\interp$ of the symbols in $F$.
We construct a sequence of interpretations $\interp = \interp_{0},
\interp_{1}, \ldots, \interp_m$, where each interpretation $\interpi$ is
generated by extending its predecessor $\interpimm$ by letting
$\jinterp = \interpimm$ and $\interpi = \jinterph$ in Lemma
\ref{G-hG-lemma} or a similar one for predicate applications.  The
effect is to include in $\interpi$ interpretations of the domain or
propositional variables introduced when eliminating the $\ordth{i}$
function or predicate symbol.  We then define interpretation
$\interps$ to be identical to $\interp_m$ for every variable appearing
in $\sF$.  By induction, we have $\einterps{\sF} = \einterp{F}$.  Since
$\sF$ is universally valid, we have $\einterp{F} = \einterps{\sF}
= \true$.  Since this construction can be performed for any
interpretation $\interp$, $F$ must also be universally valid.

{\it Only if:}  Assume $F$ is universally valid.
Starting with an interpretation $\interps$ of the domain
and propositional variables of $\sF$, we can define a sequence of
interpretations $\interps = \interp_{m}, \interp_{m-1}, \ldots,
\interp_0$, using the construction in the proof of Lemma
\ref{hG-G-lemma} (or a similar one for predicate applications) to
generate an interpretation of each function or predicate symbol in
$F$.  We then define interpretation $\interp$ to be identical to
$\interp_0$ for every function or predicate symbol appearing in $F$.
By induction, we have $\einterp{F} = \einterps{\sF}$.  Since $F$ is
universally valid, we have $\einterps{\sF} = \einterp{F} = \true$.
Since this construction can be performed for any interpretation
$\interps$, $\sF$ must also be universally valid.
\end{proof}

\subsection{Assigning Distinct Values to Variables Representing P-Function Applications}

Suppose we are given a PEUF p-formula $F$.  We can also consider this
to be a formula in EUF and hence apply the function and predicate
application elimination procedure just described to derive a formula
$\sF$ containing only domain and propositional variables.  For each
function symbol $f$ in $F$, we will introduce a series of domain
variables $\vf_1, \ldots, \vf_n$.  We will show that if $f$ is a
p-function symbol, then our decision procedure can exploit maximal
diversity by considering only interpretations that assign distinct
values to the $\vf_1, \ldots, \vf_n$.  More precisely, we need only
consider interpretations that are diverse for these variables when
deciding the validity of $F$.  This property holds even if the
variables $\vf_1, \ldots, \vf_n$ are not classified as p-function
symbols in $\sF$.

For example, consider the formula created by eliminating function
symbol $g$ from $\egF$, shown in the middle of Figure
\ref{ite-expansion-figure}.  By using an interpretation $\interps$
that assigns distinct values $1$, $2$, and $3$ to variables $\vg_1$,
$\vg_2$, and $\vg_3$ we generate distinct values for the terms $U_1$,
$U_2$, and $U_3$ (Equation \ref{variable-transform-equation}), except
when there are matches between the arguments $x$, $y$, and $\vg_1$.
On the other hand, our encoding still considers the possibility that
the arguments to the different applications of $g$ may match under
some interpretations, in which case the function results should match
as well.  Observe that the equations $\compare{x}{\vg_1}$ and
$\compare{y}{\vg_1}$ control ITEs in the transformed formula.
Nonetheless, we will show that we can prove universal validity by
considering only diverse interpretations of $\vg_1$.

To show this formally, consider the effect of replacing all instances
of a function symbol $f$ in a formula $G$ by nested $\ITE$ terms, as
described earlier, yielding a formula $\hG$ with new domain variables
$\vf_1, \ldots, \vf_n$.  We first show that when we generate these
variables while eliminating p-function applications, we can assume
they have a diverse interpretation.
\begin{lemma}
Let $\Sym$ be a subset of the symbols in $G$, and let $\hG$ be the
result of eliminating function symbol $f$ from $G$ by introducing new
domain variables $\vf_1, \ldots, \vf_n$.  If $f \in \Sym$, then for
any interpretation $\jinterp$ that is diverse for $G$ with respect to
$\Sym$, there is an interpretation $\jinterph$ that is diverse for
$\hG$ with respect to $\Sym - \{f\} \cup \{\vf_1, \ldots, \vf_n\}$
such that $\ejinterph{\hG} = \ejinterp{G}$.
\label{preserve-diverse-lemma}
\end{lemma}

\begin{proof}
Given interpretation $\jinterp$ defined over domain $\domain$, we
define interpretation $\jinterph$ over a domain $\domainh \doteq
\domain \cup \{z_1, \ldots, z_n\}$.  Each $z_i$ is a unique value,
i.e., $z_i \not = z_j$ for any $i \not = j$, and $z_i \not \in \domain$.

The proof of this lemma is based on a refinement of the proof of Lemma
\ref{G-hG-lemma}.  Whereas the construction in the earlier proof
assigned arbitrary values to the new domain variables in some cases, we
select an assignment that is diverse in these variables.
As in the construction in the proof of Lemma
\ref{G-hG-lemma}, we define $\jinterph$ for any function or predicate
symbol in $G$ to be identical to that of $\jinterp$ when the arguments
are all elements of $\domain$.  When some argument is not in
$\domain$, we let the function (respectively, predicate) application
yield an arbitrary domain (resp., truth) value.

For domain variable $\vf_i$ introduced when generating term
$U_i$, we consider two cases.  For the case where $\lma{\jinterph}{i}
= i$, we define $\ajinterph{\vf_i} = \ejinterph{T_{i}}$, i.e., the
value of the $\ordth{i}$ $f$-application term in $G$ under $\jinterp$.
For the case where $\lma{\jinterph}{i} < i$, we define
$\ajinterph{\vf_i} = z_i$.  We saw in the proof of Lemma
\ref{G-hG-lemma} that we could assign arbitrary values in this latter
case and still have $\ejinterph{\hG} = \ejinterp{G}$.  In fact, for
every subexpression $E$ of $G$, we have that its counterpart $\hE$ in
$\hG$ satisfies $\ejinterph{\hE} = \ejinterp{E}$.

We must show that $\jinterph$ is diverse for $\hG$ with respect to
$\Sym - \{f\} \cup \{\vf_1, \ldots, \vf_n\}$.  We first observe that
$\jinterph$ is identical to $\jinterp$ for all function application
terms in $G$, and hence $\jinterph$ must be diverse with respect to
$\Sym$ for $G$.  We also observe that $\jinterph$ assigns to each
variable $\vf_i$ either a unique value $z_i$ or the value yielded by
$f$-application term $T_i$ in $G$ under $\jinterph$.

Suppose there were distinct variables $\vf_i$ and $\vf_j$ such that
$\ejinterph{\vf_i} = \ejinterph{\vf_j}$.  This could occur only for the
case that $\ajinterph{\vf_i} = \ejinterph{T_i} = \ejinterph{T_j} =
\ajinterph{\vf_j}$.  Since $\jinterp$ is diverse, we can have
$\ejinterph{T_i} = \ejinterph{T_j}$ only if
$\lma{\jinterph}{i} =
\lma{\jinterph}{j}$.  We cannot have both $\lma{\jinterph}{i} = i$ and
$\lma{\jinterph}{j} = j$, and hence either $\vf_i$ or $\vf_j$ would have
been assigned unique value $z_i$ or $z_j$, respectively.  Thus, we
can conclude that $\ejinterph{\vf_i} \not = \ejinterph{\vf_j}$ for
distinct variables $\vf_i$ and $\vf_j$.

In addition, we must show that interpretation $\jinterph$ does not
create any matches between a new variable $\vf_i$ and a function
application term $T$ in $G$ that does not have $f$ as the topmost
function symbol.  Since $\jinterph$ is diverse with respect to $\Sym$
for $G$ and $f \in \Sym$, any function application term $T$ in $G$
that does not have function symbol $f$ as its topmost symbol must have
$\ejinterph{T} \not = \ejinterph{T_i}$ for all $1 \leq i \leq n$.  In
addition, we have $\ejinterph{T} \not = z_i$ for all $1 \leq i \leq
n$.  Hence, we must have $\ejinterph{T} \not = \ajinterp{\vf_i}$.
\end{proof}

We must also show that the variables introduced when eliminating
g-function applications do not adversely affect the diversity of the
other symbols.
\begin{lemma}
Let $\Sym$ be a subset of the symbols in $G$, and let $\hG$ be the
result of eliminating function symbol $f$ from $G$ by introducing new
domain variables $\vf_1, \ldots, \vf_n$.
If $f \not \in \Sym$, then for any interpretation $\jinterp$ that is diverse
for $G$ with respect to $\Sym$, there is an interpretation $\jinterph$
that is diverse for $\hG$ with respect to $\Sym$
such that $\ejinterph{\hG} = \ejinterp{G}$.
\label{protect-diverse-lemma}
\end{lemma}

\begin{proof}
The proof of this lemma is based on a refinement of the proof of Lemma
\ref{G-hG-lemma}.  Whereas the construction in the earlier proof
assigned arbitrary values to some of the new domain variables, we
select an assignment such that we do not inadvertently violate the
diversity of the other function symbols.

We define $\jinterph$ to be identical to $\jinterp$
for any symbol occurring in $G$.  For each domain variable $\vf_i$
introduced when generating term $U_i$, we define $\ajinterph{\vf_i} =
\ejinterph{T_{i}}$.  This differs from the interpretation defined in
the proof of Lemma \ref{G-hG-lemma} only in giving fixed
interpretations of domain variables that could otherwise be arbitrary,
and hence we have have $\ejinterph{\hG} = \ejinterp{G}$.  In fact, for
every subexpression $E$ of $G$, we have that its counterpart $\hE$ in
$\hG$ satisfies $\ejinterph{\hE} = \ejinterp{E}$.

We must show that $\jinterph$ is diverse for $\hG$ with respect to $\Sym$.  We
first observe that $\jinterph$ is identical to $\jinterp$ for all
function application terms in $G$, and hence $\jinterph$ must be diverse 
for $G$ with respect to $\Sym$.  We also observe that $\jinterph$ assigns
to each variable $\vf_i$ the value of $f$-application term $T_i$.  For
term $T$ having the application of function symbol $g \in \Sym$ as the
topmost operation, we must have $\ejinterph{\hT} = \ejinterph{T} \not
= \ejinterph{T_i} = \ejinterp{\vf_i}$.  Hence, we are assured that the
values assigned to the new variables under $\jinterph$ do not violate
the diversity of the interpretations of the symbols in $\Sym$.
\end{proof}

Suppose we apply the transformation process of Theorem
\ref{euf-transform-theorem} to a p-formula $F$ to generate a formula
$\sF$, and that in this process, we introduce a set of new domain
variables $V$ to replace the applications of the p-function symbols.
Let $\Symps(F)$ be the union of the set of domain variables in
$\Symp(F)$ and $V$.  That is, $\Symps(F)$ consists of those domain
variables in the original formula $F$ that were p-function symbols as
well as the domain variables generated when replacing applications of
p-function symbols.  Let $\Symgs(F)$ be the domain variables in $\sF$
that are not in $\Symps(F)$.  These variables were either g-function
symbols in $F$ or were generated when replacing g-function
applications.

We observe that we can generate all maximally
diverse interpretations of $F$ by considering only interpretations of
the variables in $\sF$ that assign distinct values to the variables in
$\Symps(F)$:
\begin{theorem}
PEUF p-formula $F$ is universally valid if and only if its translation
$\sF$ is true for every interpretation $\interps$ that is diverse over
$\Symps(F)$.
\label{distinct-theorem}
\end{theorem}

\begin{proof}
{\it Only if:}
By Theorem \ref{euf-transform-theorem}, the universal validity of $F$
implies that of $\sF$, and hence it must be true for every interpretation.

{\it If:}
The proof in the other direction follows by
inducting on the number of function and predicate symbols in $F$
having nonzero order.  For the induction step we use Lemma
\ref{preserve-diverse-lemma} when eliminating all applications of a
p-function symbol, and Lemma \ref{protect-diverse-lemma} when
eliminating all applications of a g-function symbol.  When eliminating
a predicate symbol, we do not introduce any new domain variables.
\end{proof}

\subsubsection{Discussion}

\begin{figure}
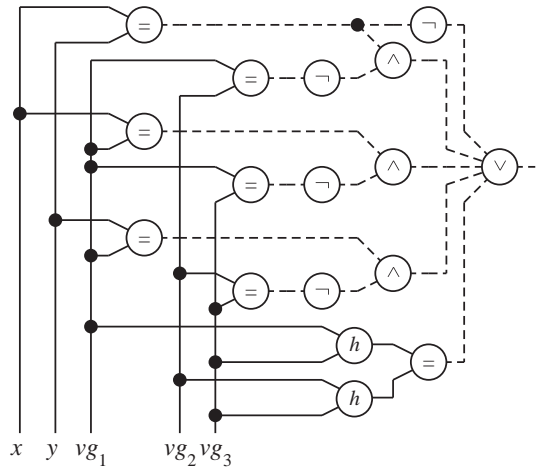
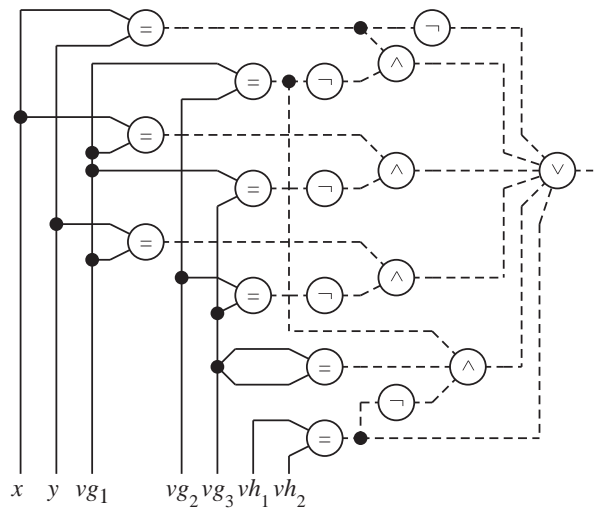

Initial formula:\\[1.5ex]
\centerline{\psfig{figure=\findfigure{eg-start},width=2.24in}}
After removing applications of function symbol $g$:\\[1.5ex]
\centerline{\psfig{figure=\findfigure{eg-ack-partial},width=2.8in}}
After removing applications of function symbol $h$:\\[1.5ex]
\centerline{\psfig{figure=\findfigure{eg-ack-final},width=3.1in}}
\caption{Ackermann's Method for Replacing Function Applications
in $\egF$.}
\label{ackermann-figure}
\end{figure}

Ackermann also describes a scheme for replacing function application
terms by domain variables \cite{ackermann-54}.  His scheme simply
replaces each instance of a function application by a newly-generated
domain variable and then introduces constraints expressing functional
consistency as antecedents to the modified formula.  As an
illustration, Figure \ref{ackermann-figure} shows the result of
applying his method to formula $\egF$ of Equation \ref{eg-formula}.
First, we replace the three applications of function symbol $g$
with new domain variables $\vg_1$, $\vg_2$, and $\vg_3$.  To maintain
functional consistency we add constraints
\begin{displaymath}
(\compare{x}{y} \Rightarrow \compare{\vg_1}{\vg_2}) \land
(\compare{x}{\vg_1} \Rightarrow \compare{\vg_1}{\vg_3}) \land
(\compare{y}{\vg_1} \Rightarrow \compare{\vg_2}{\vg_3})
\end{displaymath}
as an antecedent to the modified g-formula.  The result is shown in the
middle of Figure \ref{ackermann-figure}, using Boolean connectives
$\land$, $\lor$, and $\neg$ rather than $\Rightarrow$.  In this
diagram, the three constraints listed above form the middle three
arguments of the final disjunction.  A similar process is used to
replace the applications of function symbol $h$, adding a fourth
constraint $\compare{\vg_1}{\vg_2} \land \compare{\vg_3}{\vg_3}
\Rightarrow \compare{\vh_1}{\vh_2}$.  The result is shown at the
bottom of Figure \ref{ackermann-figure}.

There is no clear way to exploit the maximal diversity with this
translated form.  For example, if we consider only diverse
interpretations of variables $\vg_1$, $\vg_2$, and $\vg_3$, we will
fail to consider interpretations of the original g-formula for which $x$
equals $y$.

\subsection{Using Fixed Interpretations of the Variables in $\Symps(F)$}

We can further simplify the task of determining universal validity by
choosing particular domains of sufficient size and assigning fixed
interpretations to the variables in $\Symps(F)$.  The next result follows
from Theorem \ref{distinct-theorem}.

\begin{corollary}
Let $\Dp$ and $\Dg$ be
disjoint subsets of domain $\domain$ such that $|\Dp| \geq |\Symps(F)|$
and $|\Dg| \geq |\Symgs(F)|$.  Let $\alpha$ be any \onetoone{} mapping
$\alpha \colon \Symps(F) \rightarrow \Dp$.
PEUF p-formula $F$ is universally valid if and only if its translation
$\sF$ is true for every interpretation $\interps$ such that
$\ainterps{v_p} = \alpha(v_p)$ for every variable $v_p \in
\Symps(F)$, and $\ainterps{v_g} \in \Dg$ for every variable $v_g \in
\Symgs(F)$.
\label{fixed-value-corollary}
\end{corollary}

\begin{proof}
Consider any interpretation $\jinterps$ of the variables in $\Symps(F)
\cup \Symgs(F)$ that is diverse over $\Symps(F)$.  We show that we can
construct an isomorphic interpretation $\interps$ that satisfies the
restrictions of the corollary.  

Let $\Dp'$ (respectively, $\Dg'$) be
the range of $\jinterps$ considering only variables in $\Symps(F)$
(resp., $\Symgs(F)$).  The function $\jinterps \colon \Symps(F)
\rightarrow \Dp'$ must be a bijection and hence have an inverse
${\jinterps}^{-1} \colon \Dp' \rightarrow \Symps(F)$.  Furthermore, we
must have $|\Dg'| \leq |\Symgs(F)| \leq |\Dg|$.  Let $\sigma_p$ be the
\onetoone{} mapping $\sigma_p \colon \Dp' \rightarrow \Dp$ defined for
any $z$ in $\Dp'$, as $\sigma_p(z) = \alpha({\jinterps}^{-1}(z))$.
Let $\sigma_g$ be an arbitrary \onetoone{} mapping $\sigma_g \colon
\Dg' \rightarrow \Dg$.  We now define $\interps$ such that for any
variable $v$ in $\Symps(F)$ (respectively, $\Symgs(F)$) we have
$\ainterps{v}$ equal to $\sigma_p(\ajinterps{v})$ (resp.,
$\sigma_g(\ajinterps{v})$).  Finally, for any propositional variable
$a$, we let $\ainterps{a}$ equal $\ajinterps{a}$.

For any EUF formula, isomorphic interpretations will always yield
identical valuations, giving $\einterps{\sF} = \ejinterps{\sF}$.
Hence the set of interpretations satisfying the restrictions of the
corollary form a sufficient set to prove the universal validity of
$\sF$.
\end{proof}

\section{Reductions to Propositional Logic}

We present two different methods of translating a PEUF p-formula into a
propositional formula that is tautological if and only if the original
p-formula is universally valid.  Both use the function and predicate
elimination method described in the previous section so that the
translation can be applied to a formula $\sF$ containing only domain
and predicate variables.  In addition, we assume that a subset of the
domain variables $\Symps(F)$ has been identified such that we need to
encode only those interpretations that are diverse over these
variables.

\subsection{Translation Based on Bit Vector Interpretations}

A formula such as $\sF$ containing only domain and propositional
variables can readily be translated into one in propositional logic,
using the set of bit vectors of some length $k$ greater than or equal
to $\log_2 m $ as the domain of interpretation for a formula
containing $m$ domain variables \cite{velev-fmcad98}.  Domain
variables are represented with vectors of propositional variables.  In
this formulation, we represent a domain variable as a vector of
propositional variables, where truth value $\false$ encodes bit value
0, and truth value $\true$ encodes bit value 1.  In
\cite{velev-fmcad98} we described an encoding scheme in which the
$\ordth{i}$ domain variable is encoded as a bit vector of the form
$\langle 0, \ldots, 0, a_{i,k-1}, \ldots, a_{i,0} \rangle$ where $k = \lceil
\log_2 i \rceil$, and each $a_{i,j}$ is a propositional variable.
This scheme can be viewed as encoding interpretations of the domain
variables over the integers where the $\ordth{i}$ domain variable
ranges over the set $\{0, \ldots, i-1\}$ \cite{pnueli-cav99}.  That
is, it may equal any of its predecessors, or it may be distinct.

We then recursively translate $\sF$ using vectors of propositional
formulas to represent terms.  By this means we then reduce $\sF$ to a
propositional formula that is tautological if and only if $\sF$, and
consequently the original EUF formula $F$, is universally valid.

We can exploit positive equality by using fixed bit vectors, rather
than vectors of propositional variables when encoding variables in
$\Symps(F)$.  Furthermore, we can construct our bit encodings such
that the vectors encoding variables in $\Symgs(F)$ never match the bit
patterns encoding variables in $\Symps(F)$.  As an illustration,
consider formula $\egF$ given by Equation \ref{eg-formula} translated
into formula $\segF$ as diagrammed at the bottom of Figure
\ref{ite-expansion-figure}.  We need encode only those interpretations of
variables $x$, $y$, $\vg_1$, $\vg_2$, $\vg_3$, $\vh_1$, and
$\vh_2$ that are diverse respect to the last five variables.
Therefore, we can assign 3-bit encodings to the seven variables as
follows:
\begin{center}
\begin{tabular}{|c|l|}
\hline
$x$ & $\langle 0, 0, 0 \rangle$\\
$y$ & $\langle 0, 0, a_{1,0} \rangle$\\
$\vg_1$ & $\langle 0, 1, 0 \rangle$\\
$\vg_2$ & $\langle 0, 1, 1 \rangle$\\
$\vg_3$ & $\langle 1, 0, 0 \rangle$\\
$\vh_1$ & $\langle 1, 0, 1 \rangle$\\
$\vh_2$ & $\langle 1, 1, 0 \rangle$\\
\hline
\end{tabular}
\end{center}
where $a_{1,0}$ is a propositional variable.  This encoding uses the
same scheme as \cite{velev-fmcad98} for the variables in $\Symgs(F)$
but uses fixed bit patterns for the variables in $\Symps(F)$.  As a
consequence, we require just a single propositional variable to encode
formula $\segF$.

As a further refinement, we could apply methods devised by Pnueli {\em
et al.}\  to reduce the size of the domains associated with each variable
in $\Symgs(F)$ \cite{pnueli-cav99}.  This will in turn allow us to
reduce the number of propositional variables required to encode each
domain variable in $\Symgs(F)$.

\subsection{Translation Based on Pairwise Encodings of Term Equality}

Goel {\em et al.}\  \cite{goel-cav98} describe a method for generating a
propositional formula from an EUF formula, such that the propositional
formula will be a tautology if and only if the EUF formula is
universally valid.  They first use Ackermann's method to eliminate
function applications of nonzero order \cite{ackermann-54}.  Then they
introduce a propositional variable $e_{i,j}$ for each pair of domain
variables $v_i$ and $v_j$ encoding the conditions under which the two
variables have matching values.  Finally, they generate a
propositional formula in terms of the $e_{i,j}$ variables.

We provide a modified formulation of their approach that exploits the
properties of p-formulas to encode only valuations under maximally
diverse interpretations.  As a consequence, we require $e_{i,j}$
variables only to express equality among those domain variables that
represent g-term values in the original p-formula.

The propositional formula generated by either of these schemes does
not enforce constraints among the $e_{i,j}$ variables due to the
transitivity of equality, i.e., constraints of the form $e_{i,j} \land
e_{j,k} \Rightarrow e_{i,k}$.  As a result, in attempting to prove the
formula is a tautology, a false ``counterexamples'' may be generated.
We return to this issue later in this section

\subsubsection{Construction of Propositional Formula}

Starting with p-formula $F$, we apply our method of eliminating
function applications to give a formula $\sF$ containing only domain
and propositional variables.  The domain variables in $\sF$ are
partitioned into sets $\Symps(F)$, corresponding to p-function
applications in $F$, and $\Symgs(F)$ corresponding to g-function
applications in $F$.  Let us identify the variables in $\Symgs(F)$
as $\{v_1, \ldots, v_N\}$, and the variables in $\Symps(F)$ as
$\{v_{N+1}, \ldots, v_{N+M}\}$.  We need encode only those interpretations
that are diverse in this latter set of variables.

For values of $i$ and $j$ such that $1 \leq i < j \leq N$, define
propositional variables $e_{i,j}$ encoding the equality relation
between variables $v_i$ and $v_j$.  We  require these
propositional variables only for indices less than or equal to $N$.  Higher
indices correspond to variables in $\Symps(F)$, and we can assume
for any such variable $v_i$ that it will equal variable $v_j$ only
when $i=j$.

For each term $T$ in $\sF$, and each $v_i$ with $1 \leq i \leq N+M$,
we generate formulas of the form $\encodet{T}{i}$ for $1 \leq i \leq N+M$ to
encode the conditions under which the control g-formulas in the $\ITE$s
in term $T$ will be set so that value of $T$ becomes that of domain
variable $v_i$.
In addition, for each g-formula $G$ we define a propositional
formula $\encodef{G}$ giving the encoded form of $G$.  These
formulas are defined by mutual recursion.  The base cases are:
\begin{displaymath}
\begin{array}{rcll}
\encodef{\true} & \doteq & \true\\
\encodef{\false} & \doteq & \false\\
\encodef{a} & \doteq & a, & \mbox{$a$ is a propositional variable}\\
\encodet{v_i}{i} & \doteq & \true\\
\encodet{v_i}{j} & \doteq & \false, & \mbox{For $i\not = j$}\\
\end{array}
\end{displaymath}
For the logical connectives, we define ${\it encf}$ in the obvious way:
\begin{eqnarray*}
\encodef{\neg G_1} & \doteq & \neg \encodef{G_1}\\
\encodef{G_1 \land G_2} & \doteq & \encodef{G_1} \land \encodef{G_2}\\
\encodef{G_1 \lor G_2} & \doteq & \encodef{G_1} \lor \encodef{G_2}\\
\end{eqnarray*}
For $\ITE$ terms, we define ${\it enct}$ as:
\begin{eqnarray*}
\encodet{\ITE(G,T_1,T_2)}{i} & \doteq & \encodef{G} \land \encodet{T_1}{i}\;\; \lor\;\; \neg \encodef{G}
\land \encodet{T_2}{i}
\end{eqnarray*}
For equations, we define
$\encodef{\compare{T_1}{T_2}}$ to be
\begin{eqnarray}
\encodef{\compare{T_1}{T_2}} & \doteq &  
\bigvee_{1\leq i, j \leq N} \encodet{T_1}{i} \land \edgeval{i}{j} \land \encodet{T_2}{j} 
\;\;\lor\;\;  
\bigvee_{N+1\leq i \leq N+M} \encodet{T_1}{i} \land \encodet{T_2}{i} 
\nonumber \\
\label{equation-equation} 
\end{eqnarray}
where $\edgeval{i}{j}$ is defined for $1 \leq i,j \leq N$ as:
\begin{eqnarray*}
\edgeval{i}{j} & \doteq &
\left\{
\begin{array}{ll}
\true & i = j\\
e_{i,j} & i < j\\
e_{j,i} & i > j
\end{array}
\right . 
\end{eqnarray*}
Informally, 
Equation \ref{equation-equation} expresses the property that 
there are two ways for a pair of terms to be equal in an interpretation.
The first way is if the two terms evaluate to the same
variable, i.e., we have both $\encodet{T_1}{i}$ and $\encodet{T_2}{i}$ 
hold for
some variable $v_i$.  For $1 \leq i \leq N$, the left hand part of
Equation \ref{equation-equation} will hold since $\edgeval{i}{i} =
\true$.  For $N+1 \leq i \leq N$, the right hand part of Equation
\ref{equation-equation} will hold.  
The second way is that two terms will be
equal under some interpretation when they evaluate to two different
variables $v_i$ and $v_j$ that have the same value.  
In this case we will have
$\encodet{T_1}{i}$, $\encodet{T_2}{j}$, and $\edgeval{i}{j}$ hold, where $1 \leq
i,j \leq N$.  Observe that Equation \ref{equation-equation} encodes
only interpretations that are diverse over $\{v_{N+1}, \ldots,
v_{N+M}\}$.  It makes use of the fact that when $N+1 \leq i \leq N+M$,
variable $v_i$ will equal variable $v_j$ only if $i = j$.

\begin{figure}
\centerline{\psfig{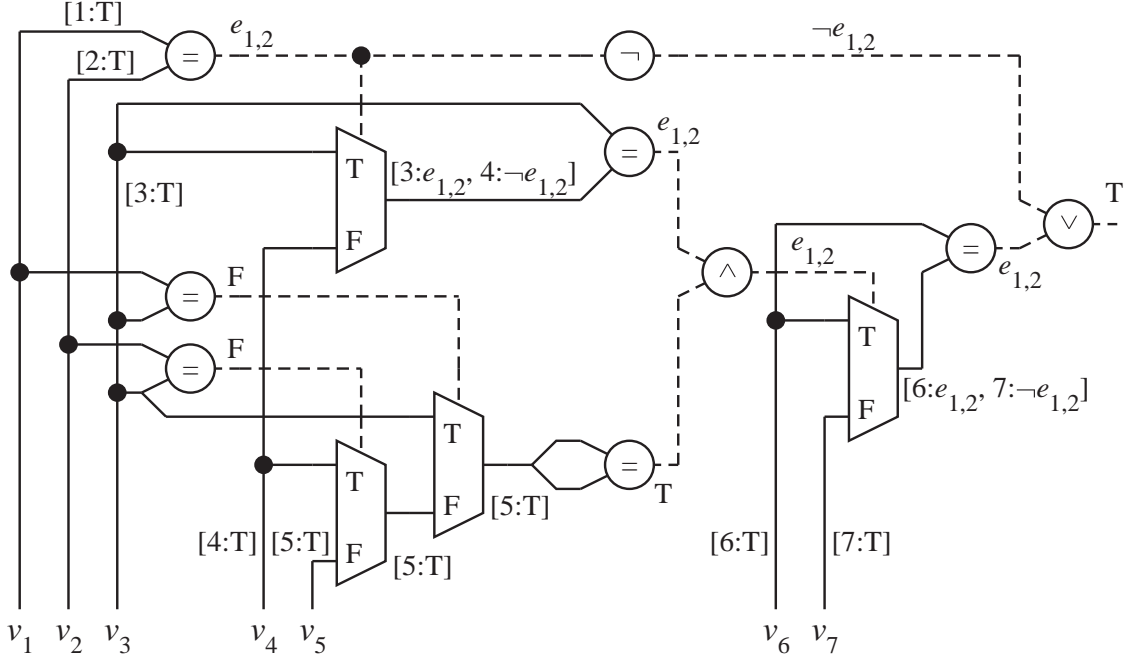}}
\caption{Encoding Example Formula in Propositional Logic. Each term $T$ is
represented as a list giving the non-$\false$ values of $\encodet{T}{i}$.}
\label{prop-figure}
\end{figure}

As an example, Figure \ref{prop-figure} shows an encoding of formula
$\sF$ given in Figure \ref{ite-expansion-figure}, which was derived
from the original formula $F$ shown in Figure \ref{eg-start-figure}.
The variables in $\Symgs(\sF)$ are $x$ and $y$.  These are renamed as
$v_1$ and $v_2$, giving $N=2$.  The variables in $\Symps(\sF)$ are
$\vg_1$, $\vg_2$, $\vg_3$, $\vh_1$, and $\vh_2$.  These are relabeled
as $v_3$ through $v_7$, giving $M=5$.  Each formula in the figure is
annotated by a (simplified) propositional formula, while each term $T$
is annotated by a list with entries of the form $i\colon
\encodet{T}{i}$, for those entries such that $\encodet{T}{i} \not =
\false$.  We use the shorthand notation ``T'' for $\true$ and ``F''
for $\false$.  Our encoding introduces a single propositional variable
$e_{1,2}$.  It can be seen that our method encodes only the
interpretations for $\sF$ labeled as D1 and D2 in Table
\ref{partition-table}.  When $e_{1,2}$ is false, we encode
interpretation D2, in which $x \not = y$ and every function
application term yields a distinct value.  When $e_{1,2}$ is true, we
encode interpretation D1, in which $x=y$ and hence we have $g(x)=g(y)$
and $h(g(x),g(g(x))) = h(g(y),g(g(y)))$.

In general, the final result of the recursive translation will be a
propositional formula $\encodef{\sF}$.  The variables in this formula
consist of the propositional variables that occur in $\sF$ as well as
a subset of the variables of the form $e_{i,j}$.  Nothing in this
formula enforces the transitivity of equality.  We will discuss in the
next section how to impose transitivity constraints in a way that
exploits the sparse structure of the equations.
Other than transitivity, we claim that the translation $\encodef{\sF}$
captures validity of $\sF$, and consequently the original p-formula
$F$.  For an interpretation $\jinterp$ over a set of propositional
variables, including variables of the form $e_{i,j}$ for $1 \leq i < j
\leq N$, we say that $\jinterp$ {\em obeys transitivity} when for all
$i$, $j$, and $k$ such that $1 \leq i, j, k \leq N$
we have $\ejinterp{\edgeval{i}{j}} \land \ejinterp{\edgeval{j}{k}} \Rightarrow
\ejinterp{\edgeval{i}{k}}$.

To formalize the intuition behind the encoding, let $\interps$ be an
interpretation of the variables in the translated formula $\sF$.  For
interpretation $\interps$, define $\select{\interps}{T}$ to be a
function mapping each term $T$ in $\sF$ to the index of the unique
domain variable selected
by the values of the $\ITE$ control g-formulas in $T$.  That
is, $\select{\interps}{v_i} \doteq i$, while
$\select{\interps}{\ITE(G,T_1,T_2)}$ is defined as
$\select{\interps}{T_1}$ when $\einterps{G} = \true$ and as
$\select{\interps}{T_2}$ when $\einterps{G} = \false$.
\begin{proposition}
\label{select-proposition}
For all interpretations $\interps$ of the variables in $\sF$ and any
term $T$ occurring in $\sF$, if $\select{\interps}{T} = i$, then
$\einterps{T} = \ainterps{v_i}$.
\end{proposition}

\begin{lemma}
For any interpretation $\interps$ of the variables in $\sF$ that is
diverse for $\Symps(F)$, there is an interpretation $\jinterp$ of the
variables in $\encodef{\sF}$ that obeys transitivity and such that
$\ejinterp{\encodef{\sF}} = \einterps{\sF}$.
\label{F-eF-lemma}
\end{lemma}

\begin{proof}
For each propositional variable $a$ occurring in $\sF$, we define
$\ajinterp{a} \doteq \ainterps{a}$.  For each pair of variables $v_i$
and $v_j$ such that $1 \leq i < j \leq N$, we define
$\ajinterp{e_{i,j}}$ to be $\true$ iff $\ainterps{v_i} =
\ainterps{v_j}$.  We can see that $\jinterp$ must obey transitivity,
because it is defined in terms of a transitive relation in $\interps$.

We prove the following  hypothesis by induction on the expression depths:
\begin{enumerate}
\item
For every formula $G$ in $\sF$: $\ejinterp{\encodef{G}} = \einterps{G}$.
\item
For every term $T$ in $\sF$ and all $i$ such that $1 \leq i \leq N+M$:
$\ejinterp{\encodet{T}{i}} = \true$ iff $\select{\interps}{T} = i$.
\end{enumerate}

The base cases hold as follows:
\begin{enumerate}
\item
Formulas of the form $\true$, $\false$, and $a$ have $\encodef{G} = G$ and
$\ejinterp{G} = \einterps{G}$.
\item
Term $v_j$ has $\ejinterp{\encodet{v_j}{i}} = \true$ iff $j=i$, and
$\select{\interps}{v_j} = i$ iff $j=i$.
\end{enumerate}

Assuming the induction hypothesis holds for formulas $G_1$ and $G_2$,
one can readily see that it will hold for formulas $\neg G_1$, $G_1
\land G_2$, and $G_1 \lor G_2$, by the definition of ${\it encf}$

Assuming the induction hypothesis holds for formula $G$ and for terms
$T_1$ and $T_2$, consider term $T$ of the form $\ITE(G,T_1,T_2)$.
For the case where $\einterps{G} = \true$, we have
$\einterps{T} = \einterps{T_1}$, and also $\select{\interps}{T} =
\select{\interps}{T_1}$.  The induction hypotheses for $T_1$ gives
$\ejinterp{\encodet{T_1}{i}} = \true$ iff $\select{\interps}{T_1} = i$.  The
induction hypothesis for $G$ gives $\ejinterp{\encodef{G}} =
\einterps{G} = \true$, and hence $\ejinterp{\encodet{T}{i}} = \ejinterp{\encodet{T_1}{i}}$.
From all this, we can conclude that $\ejinterp{\encodet{T}{i}} = \true$ iff
$\select{\interps}{T} = i$.  A similar argument holds when
$\einterps{G} = \false$, but based on the induction hypothesis for
$T_2$.

Finally, assuming the induction hypothesis holds for terms $T_1$ and
$T_2$, consider the equation $\compare{T_1}{T_2}$.  Suppose that
$\select{\interps}{T_1} = i$ and $\select{\interps}{T_2} = j$.  Our induction
hypothesis for $T_1$ and $T_2$ give
$\ejinterp{\encodet{T_1}{i}} = \ejinterp{\encodet{T_2}{j}} = \true$.
Suppose either $i > N$ or $j > N$.  Then we will have $\ainterps{v_i}
= \ainterps{v_j}$ iff $i = j$.  In addition, the right hand part of
Equation \ref{equation-equation} will hold under $\jinterp$ iff $i = j$.
Otherwise, suppose that $1 \leq i,j \leq N$.  We will have
$\ainterps{v_i} = \ainterps{v_j}$ iff $\ejinterp{\edgeval{i}{j}} =
\true$.  In addition, the left hand part of Equation
\ref{equation-equation} will hold under $\jinterp$ iff
$\ejinterp{\edgeval{i}{j}} = \true$
\end{proof}

\begin{lemma}
For every interpretation $\jinterp$ of the variables in
$\encodef{\sF}$ that obeys transitivity,
there is an interpretation $\interps$ of the
variables in $\sF$ such that $\einterp{\sF} = \ejinterp{\encodef{\sF}}$.
\label{eF-F-lemma}
\end{lemma}

\begin{figure}
\centerline{\psfig{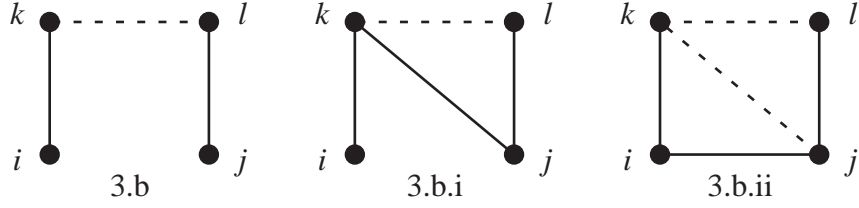}}
\caption{Case Analysis for Part 3b of Proof of Lemma
\protect{\ref{eF-F-lemma}}.
Solid lines denote equalities, while dashed lines denote inequalities.}
\label{lemma-proof-figure}
\end{figure}

\begin{proof}
We define interpretation $\interps$ over the domain of integers $\{1,
\ldots, N+M\}$.  For propositional variable $a$, we define
$\ainterps{a} = \ajinterp{a}$.  For $1 \leq j \leq N$ we let
$\ainterps{v_j}$ be the minimum value of $i$ such that
$\ejinterp{\edgeval{i}{j}} = \true$.  For $N < j \leq N+M$ we let
$\ainterps{v_j} = j$.  Observe that this interpretation gives
$\ainterps{v_j} \leq j$ for all $j \leq N$, since $\edgeval{j}{j} =
\true$, and $\ainterps{v_j} = j$ for $j > N$.

We claim that for $i \leq N$, if $\ainterps{v_j} = i$, then we must have
$\ainterps{v_i} = i$ as well.  If instead we had $\ainterps{v_i} = k <
i$, then we must have $\ejinterp{\edgeval{k}{i}} = \true$.  Combining
this with $\ejinterp{\edgeval{i}{j}} = \true$, the transitivity
requirement would give $\ejinterp{\edgeval{k}{j}} = \true$, but this
would imply that $\ainterps{v_j} = k \not = i$.  

We prove the following hypothesis by induction on the expression depths:
\begin{enumerate}
\item
For every formula $G$ in $\sF$: $\einterps{G} = \ejinterp{\encodef{G}}$.
\item
For every term $T$ in $\sF$ and all $i$ such that $1 \leq i \leq N+M$:
$\select{\interps}{T} = i$ iff $\ejinterp{\encodet{T}{i}} = \true$.
\end{enumerate}

The base cases hold as follows:
\begin{enumerate}
\item
Formulas of the form $\true$, $\false$, and $a$ have $G = \encodef{G}$ and
$\einterps{G} = \ejinterp{G}$.
\item
Term $v_j$ has $\select{\interps}{v_j} = i$ iff $j=i$ and
$\ejinterp{\encodet{v_j}{i}}  = \true$ iff $j=i$.
\end{enumerate}

Assuming the induction hypothesis holds for formula $G$ and for terms
$T_1$ and $T_2$, consider term $T$ of the form $\ITE(G,T_1,T_2)$.  For
the case where $\ejinterp{\encodef{G}} = \true$, we have
$\ejinterp{\encodet{T}{i}} = \ejinterp{\encodet{T_1}{i}}$. The induction
hypothesis for $T_1$ gives $\select{\interps}{T_1} = i$ iff
$\ejinterp{\encodet{T_1}{i}} = \true$. The induction hypothesis for $G$ gives
$\einterps{G} = \ejinterp{\encodef{G}} = \true$, giving $\einterps{T} =
\einterps{T_1}$, and also $\select{\interps}{T} =
\select{\interps}{T_1}$.  Combining all his gives
$\select{\interps}{T} = i$ iff $\ejinterp{\encodet{T}{i}} = \true$.  A
similar argument can be made when $\ejinterp{\encodef{G}} = \false$,
but based on the induction hypothesis for $T_2$.

Finally, assuming the induction hypothesis holds for terms $T_1$ and
$T_2$, consider the equation $\compare{T_1}{T_2}$.  Let $i =
\select{\interps}{T_1}$ and $j = \select{\interps}{T_2}$. In addition,
let $k = \ainterps{v_i}$ and $l = \ainterps{v_j}$.  Our induction
hypothesis gives $\ejinterp{\encodet{T_1}{i}} = \true$, and
$\ejinterp{\encodet{T_2}{j}} = \true$.  Proposition \ref{select-proposition}
gives $\einterps{T_1} = k$ and $\einterps{T_2} = l$.  By our earlier
argument, we must also have $\ainterps{v_k} = k$ and $\ainterps{v_l} =
l$.  We consider different cases for the values of $i$, $j$, $k$, and
$l$.

\begin{enumerate}
\item Suppose $i > N$.  Then we must have $k = \ainterps{v_i} = i$.
Equation $\compare{T_1}{T_2}$ will hold under $\interps$ iff
$\ainterps{v_j} = l = k$, and this will hold iff $j = l = k = i$.  In
addition, the right hand part of Equation \ref{equation-equation} will
hold under $\jinterp$ iff $i = j$.

\item Suppose $j > N$.  By an argument similar to the previous one, we
will have equation $\compare{T_1}{T_2}$ holding under interpretation
$\interps$ and Equation \ref{equation-equation} holding under
interpretation $\jinterp$ iff $i = j$.

\item
Suppose $1 \leq i,j \leq N$.  Since $\ainterps{v_i} = k =
\ainterps{v_k}$ we must have $\ejinterp{\edgeval{k}{i}} = \true$.
Similarly, since $\ainterps{v_j} = l = \ainterps{v_l}$ we must have
$\ejinterp{\edgeval{l}{j}} = \true$.  
\begin{enumerate}
\item
Suppose $k = l$, and hence $\compare{T_1}{T_2}$ holds under $\interps$.
Then we have $\ejinterp{\edgeval{i}{k}} = \ejinterp{\edgeval{k}{j}} =
\true$.  Our transitivity requirement then gives
$\ejinterp{\edgeval{i}{j}} = \true$, and hence the left hand part of
Equation \ref{equation-equation} will hold under $\jinterp$.
\item
Suppose $k \not = l$, and hence $\compare{T_1}{T_2}$ does not hold
 under $\interps$.  We must have $\ejinterp{\edgeval{k}{l}} = \false$.
 This condition is illustrated in the left hand diagram of Figure
 \ref{lemma-proof-figure}.  In this figure we use solid lines to
 denote equalities and dashed lines to denote inequalities.
We argue that we must also have 
$\ejinterp{\edgeval{i}{j}} = \false$ by the following case analysis for
$\edgeval{k}{j}$:
\begin{enumerate}
\item For $\ejinterp{\edgeval{k}{j}} = \true$, we get the case
 diagrammed in the middle of Figure \ref{lemma-proof-figure} where the
 diagonal line creates a triangle with just
 one dashed line (inequality).  This represents a violation of our
 transitivity requirement, since it indicates
 $\ejinterp{\edgeval{k}{j}} = \ejinterp{\edgeval{j}{l}} = \true$, but
 $\ejinterp{\edgeval{k}{l}} = \false$.
\item For $\ejinterp{\edgeval{k}{j}} = \false$ and
$\ejinterp{\edgeval{i}{j}} = \true$, we have the case diagrammed on
the right side of Figure \ref{lemma-proof-figure}.  Again we have a
triangle with just one dashed line indicating a violation of
our transitivity requirement, with $\ejinterp{\edgeval{k}{i}} =
\ejinterp{\edgeval{i}{j}} = \true$, but $\ejinterp{\edgeval{k}{j}} =
\false$.
\end{enumerate}
With $\ejinterp{\edgeval{i}{j}} = \false$, Equation
\ref{equation-equation} will not hold under $\jinterp$.
\end{enumerate}
\end{enumerate}
From this case analysis we see that $\compare{T_1}{T_2}$ holds under
$\interps$ iff Equation \ref{equation-equation} holds under
$\jinterp$.
\end{proof}

\begin{theorem}
p-formula $F$ is universally valid iff its translation $\encodef{\sF}$ is
true for all interpretations that obey transitivity.
\end{theorem}

\begin{proof}
This theorem follows directly from Lemmas \ref{F-eF-lemma} and
\ref{eF-F-lemma}.
\end{proof}

We have thus reduced the task of proving that a PEUF p-formula is
universally valid to one of proving that a propositional formula is
true under all interpretations that satisfy transitivity constraints.
This result is similar to that of Goel {\em et al.}, except that they
potentially require a propositional variable for every pair of
function application terms occurring in the original formula.  In our
case, we only introduce these variables for a subset of the pairs of
g-function applications.  For example, their method would require 8
variables to encode the transformed version of formula $\egF$ shown in
Figure \ref{ackermann-figure}, whereas we require only one using
either of our two encoding schemes.

To complete the implementation of a decision procedure for PEUF, we
must devise a procedure for the {\em constrained Boolean
satisfiability} problem defined by Goel, {\em et al.}, as follows.  We
are given a Boolean formula $\fsat$ over a set of propositional variables.
A subset of the variables are
of the form $e_{i,j}$, where $1 \leq i < j \leq N$.  A {\em
transitivity constraint} is a formula of the form
\begin{eqnarray*}
e_{[i_1,i_2]}
\land e_{[i_2,i_3]} \land \cdots \land e_{[i_{k-1},i_{k}]} &
\Rightarrow &
e_{[i_1,i_k]}
\end{eqnarray*}
where $e_{[i,j]}$ equals $e_{i,j}$ when $i < j$ and equals $e_{j,i}$
when $i > j$.  The task is to find a truth assignment that satisfies
$\fsat$, as well as every transitivity constraint.  For PEUF p-formula
$F$, if we can show that the g-formula $\neg \encodef{\sF}$ has no
satisfying assignment that also satisfies the transitivity
constraints, then we have proved that $F$ is universally valid.

Goel, {\em et al.}, have shown the constrained Boolean satisfiability
problem is NP-hard, even when $\fsat$ is represented as an OBDD\@.  We
have also studied this problem in the context of pipelined processor
verification \cite{bryant-cav00,bryant-transitivity00}.  We have found
that we can exploit the sparse structure of the $e_{i,j}$ variables
both when using OBDDs to perform the verification and when using
Boolean satisfiability checkers.  As a result, enforcing transitivity
constraints has a relatively small impact on the performance of the
decision procedure.  In fact, many processors can be verified without
considering transitivity constraints---the formula $\neg
\encodef{\sF}$ is unsatisfiable even disregarding transitivity
constraints \cite{velev-charme99}.

\section{Modeling Microprocessors in PEUF}

Our interest is in verifying pipelined microprocessors, proving their
equivalence to an unpipelined instruction set architecture model.
We use the approach pioneered by Burch and Dill \cite{burch-cav94} in
which the abstraction function from pipeline state to architectural
state is computed by symbolically simulating a flushing of the
pipeline state and then projecting away the state of all but the
architectural state elements, such as the register file, program
counter, and data memory.  Operationally, we construct two sets of
p-terms describing the final values of the state elements resulting
from two different symbolic simulation sequences---one from the
pipeline model and one from the instruction set model.  The
correctness condition is represented by a p-formula expressing the
equality of these two sets of p-terms.

Our approach starts with an RTL or gate-level model of the
microprocessor and performs a series of abstractions to create a model
of the data path using terms that satisfy
the restrictions of PEUF{}.  Examining the structure of a pipelined
processor, we find that the signals we wish to abstract as terms can
be classified as follows:
\begin{description}
\item{{\bf Program Data:}}
Values generated by the ALU and stored in registers and
data memory.  These are also used as addresses for the data memory.
\item{{\bf Register Identifiers:}} Used to index the register file
\item{{\bf Instruction Addresses:}} Used to designate which instructions to fetch
\item{{\bf Control values:}} Status flags, opcodes, and other
signals modeled at the bit level.
\end{description}
By proper construction of
the data path model, both program data and instruction addresses can
be represented as p-terms.  Register identifiers, on the other hand, must be
modeled as g-terms, because their comparisons control the stall and
bypass logic.  The remaining control logic is kept at the bit level.

In order to generate such a model, we must abstract the operation of
some of the processor units.  For example, the data path ALU is
abstracted as an uninterpreted p-function, generating a data value
given its data and control inputs.
Formally, this requires extending
the syntax for function applications to allow both formula and term
inputs.
We model the PC incrementer and the branch target logic as
uninterpreted functions generating instruction addresses.  We model
the branch decision logic as an uninterpreted predicate indicating
whether or not to take the branch based on data and control inputs.
This allows us to abstract away the data equality test used by the
branch-on-equal instruction.

\comment{
To model the register file, we use the memory model described by Burch
and Dill \cite{burch-cav94}, creating a nested $\ITE$ structure to
record the history of writes to the memory.  
This approach requires
equations between memory addresses controlling the $\ITE$ operations.
For the register file, such equations are allowed since g-term register
identifiers serve as addresses.  For the data memory, however, the
memory addresses are g-term program data and hence such equations
cannot be used.  Instead, we model the data memory as a generic state
machine, changing state in some arbitrary way for each write
operation, and returning some arbitrary value dependent on the state
and the address for each read operation.  Such an abstraction
technique is sound, but it does not capture all of the properties of a
memory.  It is satisfactory for modeling processors in which there is
no reordering of writes relative to each other or relative to reads.
}

To model the register file, we use the memory model described by Burch
and Dill \cite{burch-cav94}, creating a nested $\ITE$ structure to
encode the effect of a read operation based on the history of writes to
the memory. That is, suppose at some point we have performed $k$ write
operations with addresses given by  terms $A_1, \ldots, A_k$ and data
given by terms $D_1, \ldots, D_k$.  Then the effect of a read with
address term $A$ is a the term:
\begin{equation}
\ITE(\compare{A}{A_k}, D_k, \ITE(\compare{A}{A_{k-1}}, D_{k-1}, \cdots
\ITE(\compare{A}{A_1}, D_1, f_I(A)) \cdots))
\label{read-equation}
\end{equation}
where $f_I$ is an uninterpreted function expressing the initial memory
state.
Note that the presence of these comparison and $\ITE$ operations requires
register identifiers to be modeled with g-terms.

Since we view the instruction memory as being read-only, we can model
the instruction memory as a collection of uninterpreted functions and
predicates---each generating a different portion of the instruction
field.  Some of these will be p-functions (for generating immediate
data), some will be g-functions (for generating register identifiers),
and some will be predicates (for generating the different bits of the
opcode).  In practice, the interpretation of different portions of an
instruction word depends on the instruction type, essentially forming a
``tagged union'' data type.  Extracting and interpreting the different
instruction fields during processor verification is an interesting
research problem, but it lies outside the scope of this paper.

The data memory provides a greater modeling challenge.  Since the
memory addresses are generated by the ALU, they are considered program
data, which we would like to model as p-terms.  However, using a
memory model similar to that used for the register file requires
comparisons between addresses and $\ITE$ operations having the
comparison results as control.  Instead, we must create a more
abstract memory model that weakens the semantics of a true memory to
satisfy the restrictions of PEUF{}.  Our abstraction models a memory
as a generic state machine, computing a new state for each write
operation based on the input data, address, and current state.  Rather
than Equation \ref{read-equation}, we would express the effect of a
read with address term $A$ after $k$ write operations as $f_r(S_k,
A)$, where $f_r$ is an uninterpreted ``memory read'' function, and
$S_k$ is a term representing the state of the memory after the $k$
write operations.  This term is defined recursively as $S_0 = s_0$,
where $s_0$ is a domain variable representing the initial state, and
$S_i = f_u(S_{i-1}, A_i, D_i)$ for $i \geq 1$, where $f_u$ is an uninterpreted
``memory update'' function.
In essence, we view write operations as making arbitrary changes to
the entire memory state.

This model removes some of the
correlations guaranteed by the read operations of an actual memory.
For example, although it will yield identical operations for two
successive read operations to the same address, it will indicate that
possibly different result could be returned if these two reads are
separated by a write, even to a different address.  In addition, if we
write data $D$ to address $A$ and then immediately read from this
address, our model will not indicate that the resulting value must be
$D$.  Nonetheless, it can readily be seen that this abstraction is a
conservative approximation of an actual memory.  As long as the
pipelined processor performs only the write operations indicated by
the program, that it performs writes in program order, and that the
ordering of reads relative to writes matches the program order, the
two simulations will produce equal terms representing the final memory
states.

The remaining parts of the data path include comparators comparing for
matching register identifiers to determine bypass and stall conditions, and
multiplexors, modeled as $\ITE$ operations selecting between alternate
data and instruction address sources.  Since register identifiers are modeled
as g-terms, these comparison and control combinations obey the
restrictions of PEUF{}.  Finally, such operations as instruction
decoding and pipeline control are modeled at the bit level using
Boolean operations.

\section{Experimental Results}

In \cite{velev-fmcad98}, we described the implementation of a symbolic
simulator for verifying pipelined systems using vectors of Boolean
variables to encode domain variables, effectively treating all terms
as g-terms.  This simulation is performed directly on a modified
gate-level representation of the processor.  In this modified version,
we replace all state holding elements (registers, memories, and
latches) with behavioral models we call Efficient Memory Models
(EMMs).  In addition all data-transformation elements (e.g., ALUs,
shifters, PC incrementers) are replaced by read-only EMMs, which
effectively implement the transformation of function applications into
nested $\ITE$ expressions described in Section
\ref{ite-expansion-section}.  One interesting feature of this
implementation is that our decision procedure is executed directly as
part of the symbolic simulation.  Whereas other implementations,
including Burch and Dill's, first generate a formula and then decide
its validity, our implementation generates and manipulates bit-vector
representations of terms as the symbolic simulation proceeds.
Modifying this program to exploit positive equality simply involves
having the EMMs generate expressions containing fixed bit patterns
rather than vectors of Boolean variables.  All performance results
presented here were measured on a 125 MHz Sun Microsystems SPARC-20.

We constructed several simple pipeline processor design based on the
MIPS instruction set \cite{kane-92}.
We abstract register
identifiers as g-terms, and hence our verification covers all possible
numbers of program registers including the 32 of the MIPS instruction
set.  The simplest version of the pipeline implements ten different
Register-Register and Register-Immediate instructions.  Our program
could verify this design in 48 seconds of CPU time and just 7 MB of
memory using vectors of Boolean variables to encode domain variables.
Using fixed bit patterns reduces the complexity of the verification to
6 seconds and 2 MB{}.

We then added a memory stage to implement load and store instructions.
An interlock stalls the processor one cycle when a load instruction is
followed by an instruction requiring the loaded result.  Treating all
terms as g-terms and using vectors of Boolean variables to encode
domain variables, we could not verify even a 4-bit version of this
data path (effectively reducing $|\domain|$ to 16), despite running
for over 2000 seconds.  The fact that both addresses and data for the
memory come from the register file induces a circular constraint on
the ordering of BDD variables encoding the terms.  On the other hand,
exploiting positive equality by using fixed bit patterns for register
values eliminates these variable ordering concerns.  As a consequence,
we could verify this design in just 12 CPU
seconds using 1.8 MB{}.

Finally, we verified a complete CPU, with a 5-stage pipeline
implementing 10 ALU instructions, load and store, and MIPS
instructions \verb@j@ (jump with target computed from instruction
word), \verb@jr@ (jump using register value as target), and \verb@beq@
(branch on equal).  This design is comparable to the DLX design
\cite{hennessy-96} verified by Burch and Dill in \cite{burch-cav94},
although our version contains more of the implementation details.
We were unable to verify this processor using the scheme of
\cite{velev-fmcad98}.  Having instruction addresses dependent on
instruction or data values leads to exponential BDD growth when
modeling the instruction memory. Modeling instruction addresses as
p-terms, on the other hand, makes this verification tractable.  We can
verify the full, 32-bit version of the processor using 169 CPU seconds
and 7.5 MB{}.

More recently \cite{velev-charme99}, we have implemented a new
decision procedure using the pairwise encoding of term equality
approach.  Verifying a single-issue RISC pipeline with this decision
procedure requires only a fraction of a CPU second.  We have been able
to verify a dual-issue pipeline with just 35 seconds of CPU time.  By
contrast, Burch \cite{burch-dac96} verified a somewhat simpler
dual-issue processor only after devising 3 different commutative
diagrams, providing 28 manual case splits, and using around 30 minutes
of CPU time.  Our results are far better than any others achieved to
date.  In more recent work \cite{velev-dac00}, we have been able to
add additional features to our pipeline model, including exception
handling, multicycle instructions, and branch prediction.  By using
appropriate abstractions, most of this complexity comes can be
expressed by p-function applications and by predicate applications.
We have also been able to verify models of VLIW processors
\cite{velev-cav00}.  These models are far more beyond the capability
of any other automated tool for verifying pipelined microprocessors.
Having a decision procedure that exploits positive equality is
critical to the success of this verifier.

\section{Conclusions}

Eliminating Boolean variables in the encoding of terms representing
program data and instruction addresses has given us a major
breakthrough in our ability to verify pipelined processors.  Our BDD
variables now encode only control conditions and register identifiers.  For
classic RISC pipelines, the resulting state space is small and regular
enough to be handled readily with BDDs.  

We believe that there are many optimizations that will yield further
improvements in the performance of Boolean methods for deciding
formulas involving uninterpreted functions.  We have found that
relaxing functional consistency constraints to allow independent
functionality of different instructions, as was done in
\cite{damm-concur98}, can dramatically improve both memory and time
performance.  We look forward to testing our scheme
for generating a propositional formula using Boolean
variables to encode the relations between terms.
Our method exploits positive equality to greatly reduce the number of
propositional variables in the generated formula,
as well as the number of functional
consistency and transitivity constraints.
We are also considering the use of satisfiability checkers rather than
BDDs for performing our tautology checking

We consider pipelined processor verification to be a ``grand
challenge'' problem for formal verification.  We have found that
complexity grows rapidly as we move to more complex pipelines,
including ones with out-of-order execution and register renaming.
Further breakthroughs will be required before we can handle complete
models of state-of-the art processors.

\end{document}